\documentclass[a4paper,11pt]{article}

\pdfoutput=1
\usepackage{amsmath}
\usepackage{graphicx}
\usepackage{amsfonts}
\usepackage{amssymb}
\usepackage{amsmath}
\usepackage{amsthm}
\usepackage{dsfont}
\usepackage{mathtools}
\usepackage{empheq}
\usepackage{bm}
\usepackage[svgnames]{xcolor}
\usepackage[colorlinks=true,urlcolor=purple,linkcolor=bleu-lpens,citecolor=magenta,bookmarks=false]{hyperref}
\usepackage{stmaryrd}
\usepackage[nomessages]{fp}
\usepackage{color}
\usepackage{empheq}

\definecolor{orange-lpens}{RGB}{230,62,37}
\definecolor{bleu-lpens}{RGB}{41,50,117}

\newcommand{\bc}{\begin{center}}
\newcommand{\ec}{\end{center}}
\newcommand{\beq}{\begin{equation}}
\newcommand{\eeq}{\end{equation}}
\newcommand{\beqa}{\begin{eqnarray}}
\newcommand{\eeqa}{\end{eqnarray}}
\newcommand{\beqs}{\begin{eqnarray*}}
\newcommand{\eeqs}{\end{eqnarray*}}

\newcommand{\bi}{\begin{itemize}}
\newcommand{\ei}{\end{itemize}}
\newcommand{\oo}{\text{\o}}

\newcommand{\bra}{\langle}
\newcommand{\ket}{\rangle}
\newcommand{\Tr}{{\rm Tr}}

\def\ket#1{|#1\rangle}
\def\bra#1{\langle#1|}
\def\vev#1{\langle #1\rangle}
\newcommand{\fn}{ \mathfrak{n} }
\newcommand{\fj}{ \mathfrak{j} }
\def\cadremath#1{\vbox{\hrule\hbox{\vrule\kern8pt\vbox{\kern8pt
			\hbox{ {$\displaystyle #1 $ } }\kern8pt} 
			\kern8pt\vrule}\hrule}}
\title{\bf Can the Macroscopic Fluctuation Theory be Quantized ?}
\author{}
\date{}

\begin{document}

 {\color{bleu-lpens}
\maketitle
\pagestyle{empty}
\vskip -1.0 truecm

\centerline{Denis BERNARD$~{}^{\clubsuit}$\footnote{denis.bernard@ens.fr}}

\vskip 0.5 truecm

{\centering{\small{$^\clubsuit$ Laboratoire de Physique de l'Ecole Normale Sup\'erieure, CNRS, ENS \& Universit\'e PSL, Sorbonne Universit\'e, Universit\'e de Paris, 75005 Paris, France.}}}
\vskip 0.5 truecm

}
\pagestyle{plain}

\begin{abstract}
The Macroscopic Fluctuation Theory is an effective framework to describe transports and their fluctuations in classical out-of-equilibrium diffusive systems. Whether the Macroscopic Fluctuation Theory may be extended to the quantum realm and which form this extension may take is yet terra incognita but is a timely question.  In this short introductory review, I discuss possible questions that a quantum version of the Macroscopic Fluctuation Theory could address and how analysing Quantum Simple Exclusion Processes yields pieces of answers to these questions. 
\end{abstract}

\vskip 0.5 truecm

{\setlength{\parskip}{0pt plus 1pt} \tableofcontents}

\medskip



\section{Introduction}

Non-equilibrium phenomena, classical or quantum, are ubiquitous in Nature, but their understanding is more delicate, and thus yet less profound, than the equilibrium ones. Last decades has witnessed important conceptual and practical progresses in this direction, both for classical and quantum systems. 
For classical systems, these progresses started with the study of fluctuating hydrodynamics of many particle systems \cite{Masi1991MathematicalMF,Kipnis1999Scaling,Liggett1999Stochastic,Spohn1991Large,Spohn2014Nonlinear,Schutz2000Exactly,Kipnis1989Hydrodynamics}, or with exact analysis of simple model systems, such as the Symmetric  or Asymetric Simple Exclusion Process (SSEP or ASEP) ~\cite{Derrida1998ASEP,Derrida2007Non-equilibrium,Mallick2015The}. They advanced with the understanding of fluctuation relations~\cite{Gallavotti1995Dynamical,Jarzynski1997Nonequilibrium,Crooks1999Entropy} and their interplay with time reversal~\cite{Maes1999The,Maes2003time}, and culminated with the formulation of the Macroscopic Fluctuation Theory (MFT) which is an effective theory describing transports and their fluctuations in diffusive classical systems~\cite{Bertini2005current,Bertini2015MFT}. 
For quantum systems, a good understanding of entanglement dynamics \cite{Calabrese2005Evolution,Bardarson2012Unbounded,Ho2017Entanglement,znidaric2020entanglement} and quenched dynamics \cite{calabrese2016quantum,essler2016quench,caux2016quench} in critical or integrable models has been obtained. Substantial information on transport has been gained for integrable many-body systems via the formulation of the Generalized Hydrodynamics (GHD) \cite{Castro2016emergent,Bertini2016Transport,Doyon_2020_GHDReview}. The latter is however well adapted to describe transport but less fitted to elucidate quantum interference phenomena. Furthermore, transports in these systems are mainly ballistic. 

The questions whether Macroscopic Fluctuation Theory (MFT) may be extended to deal with quantum systems and which form this extension would take are still open. If such theory can be formulated, it should aim at describing not only diffusive transports and their fluctuations but also quantum coherent phenomena, say the dynamics of quantum interferences or entanglement spreading and their fluctuations, in out-of-equilibrium extended, diffusive, systems.  

Pieces of information on a possible form of such theory has recently been obtained by studying model systems based on random quantum circuits \cite{Brown2015decoupling,Brandao2016local,Nahum2017Quantum,Nahum2018operator,Chan2018Solution,jonay2018coarse} for which a membrane picture \cite{Mezei2018membrane,Zhou2019emergent,Gullans2019entanglement,Agon2019bit,Zhou2020entanglement} for entanglement production in many-body systems is emerging.

In parallel, stochastic dynamics in quantum extended systems such as noisy spin chains were analysed \cite{Bauer2017Stochastic,Knap2018Entanglement,Rowlands2018noisy,frassek2020duality}. This leads us to introduce the Quantum Simple Exclusion Processes. The closed Quantum Symmetric Simple Exclusion Process (Q-SSEP), was introduced  in \cite{Bauer2019Equilibrium} (under another name), its open version in \cite{Bernard2019Open}, and its asymmetric analogue, the Quantum Asymmetric Simple Exclusion Process (Q-ASEP), was proposed in \cite{Jin2020Stochastic}. Through decoherence, the average dynamics in such models typically reduces to classical simple exclusion processes. However, this quasi-classical reduction only applies to the average dynamics. Fluctuations are beyond this quasi-classical regime and survive decoherence. They reveal patterns which are possibly generic for mesoscopic diffusive quantum systems. 

In this review, we first introduce basic notions about the MFT and pose a few questions relevant to its extension to the quantum regime and, second, discuss possible routes for exploring quantum extensions of the MFT and how these explorations lead to consider stochastic dynamics in quantum many-body systems. While classical stochastic dynamics has been a well studied domain for many decades, stochastic dynamics in quantum extended systems remained largely unexplored, to the best of our knowledge. We then present a brief introduction to the Quantum Simple Exclusion Processes and to some of the recently obtained results concerning their on-going analysis. We have chosen to select a few claims, summarising these analysis in a necessarily biased ways, but allowing for a comparison with classical MFT. For each topic, we also tried to formulate a few questions left opened by these analysis.

As a side comment, we pose a question (or propose a conjecture) about ergodicity property of monitored extended quantum systems.

\section{Why, What, and How~?}

\subsection{A glimpse on classical Macroscopic Fluctuation Theory}

\begin{figure}\bc
\includegraphics[width=0.65\textwidth]{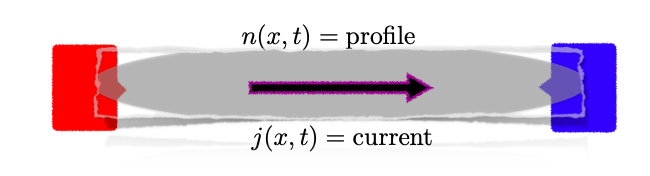}\ec
\caption{\it Classical setup to drive a diffusive systems out-of-equilibrium (the two blue/red square represent two reservoirs at different chemical potentials or temperatures).}
\label{fig:classical-setup}
\end{figure}

The Macroscopic Fluctuation Theory \cite{Bertini2005current,Bertini2015MFT} is an effective theory describing current and density fluctuations in classical out-of-equilibrium systems. It applies to locally diffusive systems satisfying the Fourier-Fick's law which asserts that the current is proportional to the density gradient. The proportionality coefficient is called the diffusion constant. Whenever the current does not vanish -- a situation which reflects transport --, the physical system is out-of-equilibrium. By the Fourier-Fick's law, diffusive systems can be maintained out-of-equilibrium by imposing a non-zero density gradient or by applying an external field. A standard set-up consists in putting the system in contact with two reservoirs at different chemical potentials\footnote{The discussion is here presented in terms of particle density and its current, but one may change the point and discuss about transport of other physical quantities, say heat or temperature, simply by changing the names.}. See Figure \ref{fig:classical-setup}.

One of the  formulation of MFT~\cite{Bertini2005current,Bertini2015MFT} starts with stochastic differential equations (SDE) for the density $\fn(x,t)$ and the current $\fj(x,t)$. In one dimension, they take the following form~: 
\begin{subequations}\label{eq:mft}
\begin{align}
& \partial_t \fn(x,t) + \partial_x \fj(x,t) = 0 ~,\label{eq:mft-a}\\
& \fj(x,t) = -D(\fn) \partial_x \fn(x,t) + \varepsilon^{1/2}\, \sigma(\fn)^{1/2}\, \xi(x,t) ~,\label{eq:mft-b}
\end{align}
\end{subequations}
with $D(\fn)$ the diffusion constant, $\sigma(\fn)$ the mobility and $\xi(x,t)$ a Gaussian space-time white noise, $\mathbb{E}[\xi(x,t)\xi(x',t')]=\delta(x-x')\delta(t-t')$. Here, $\mathbb{E}[\cdot]$ refers to the statistical average and  $\varepsilon := a_0/L$ with $a_0$ the short microscopic scale (an ultra-violet cut-off) and $L$ the linear size of the system. In particular, $\varepsilon\to 0$ in the large system size limit or in the scaling limit $a_0\to 0$ at $L$ fixed.  

The first equation \eqref{eq:mft-a} is a conservation law. The second one \eqref{eq:mft-b} is a noisy version of the Fourier-Fick's law  $\fj(x,t) = -D(\fn) \partial_x \fn(x,t)$. The mobility quantifies the response of the system to an external field $E$ which, in the linear approximation, produces an extra current $\fj_\mathrm{ext}= \sigma(\fn)\partial_x E$. It also codes for the density fluctuations at equilibrium. Both the diffusion constant and the mobility may depend non-linearly on the local density. They satisfy the Einstein relation $D(\fn)=\sigma(\fn)\, f''(\fn)$ with $f(\fn)$ the equilibrium free energy per unit of volume.

The strength of the noise in eq.\eqref{eq:mft} gets smaller as the system size increases (or as the ultra-violet cut-off is shrinked), so that analysing eq.\eqref{eq:mft} is actually a small noise problem \cite{FW2012RandomPert}. This reflects the physical fact that fluctuations in thermodynamical systems are small, sub-leading in the system size. The MFT describes rare fluctuations in out-of-equilibrium macroscopic systems.

Through eq.\eqref{eq:mft}, the statistical distribution of the noise induces that of the current and density profiles. By integrating out the noise, the probability distribution can (formally) be represented as a path integral with weight,
\beqa \label{eq:mft-weight}
\exp\Big(-\frac{1}{\varepsilon}\int \!\! dxdt\, \frac{\big(\fj(x,t) + D(\fn(x,t)) \partial_x \fn(x,t)\big)^2}{2\sigma(\fn(x,t))} \Big) ~,
\eeqa
conditioned by the conservation law $\partial_t \fn(x,t) + \partial_x \fj(x,t) = 0$. In some cases, this weight can be deduced from a simple, but elegant, additivity principle \cite{Bodineau2004Current}. Configurations extremizing this weight are the most probable. They correspond to the dominating non-fluctuating solutions of the Fourier-Fick's law depending on boundary conditions specific to the physical set-up. 

Fluctuations around the dominating configurations are sub-leading in the system size.  They satisfy a large deviation principle. The large prefactor $\varepsilon^{-1}=L/a_0$ in the weight \eqref{eq:mft-weight} echoes the weakness of the noise in eq.\eqref{eq:mft} for macroscopically large systems. It ensures that the large deviation functions are computable through solutions of extremization problems (which may nevertheless be difficult to solve) \cite{Bertini2015MFT,FW2012RandomPert,Bodineau2004Current}. 

The MFT also emerged from studies of simple representative toy models. The iconic exemple is the so-called symmetric simple exclusion process (SSEP). In its one dimensional version, it describes particles moving along a chain\footnote{Variants of SSEP can be defined on any graph.} with the exclusion that any given site cannot be occupied by more than one particle at each instant. Particles are injected and extracted at the two ends of the system interval. In the symmetric case the probability to move toward the left or the right are equal (they are different in the asymmetric case). Any configuration is specified by its occupation numbers $\mathfrak{n}_j$ at each site $j$, with $\mathfrak{n}_j=0$ for an empty site (represented as $\text{\o}$) and $\mathfrak{n}_j=1$ for a full site (represented as $\bullet$). The evolution from one configuration to another is a Markovian random process, with probability transition per unit of time given by the following Markov transition matrix~: 
\beqa \label{eq:P-ssep-classic}
\mathbb{M}_{{\rm ssep}}(|\oo \oo|) & = & 0 ~,\nonumber\\
\mathbb{M}_{{\rm ssep}}(|\!\bullet\! \oo|) & = & -|\!\bullet\! \oo|+|\oo\! \bullet\!| ~,\\
\mathbb{M}_{{\rm ssep}}(|\oo\! \bullet\!|) & = & -|\oo\! \bullet\!|+|\!\bullet \!\oo| ~, \nonumber\\
\mathbb{M}_{{\rm ssep}}(|\!\bullet \!\bullet |) & = & 0 ~.\nonumber
\eeqa
The first and last equations simply express the absence of move either because there is no particle or because  the exclusion freezes the dynamics. The two middle ones code for symmetric moves to the right or the left when the configurations allow. The evolution rules are modified at the boundaries to reflect the rates at which particles are injected and extracted. This system is known to be integrable and its steady states represented by matrix product states. There is a vast literature on exact results for SSEP, see e.g. \cite{Derrida1993exact,deGier2005Bethe,Blythe2007Nonequilibrium,Mallick2011some}. 

The large deviation function for density fluctuations in SSEP were computed from its microscopic definition \cite{Derrida2001Free,Derrida2002Large}. Let us consider a protocol as in Figure \ref{fig:classical-setup} in which one fixes the density at the two ends of the interval $[0,L]$, say $n_a$ and $n_b$. If $n_a\not= n_b$ the system is out-of-equilibrium, with a non-zero mean current proportional to the density difference $\Delta n := n_b-n_a$.
In the steady state, the density statistics satisfies a large deviation principle in the sense that the distribution of the density profiles behaves as follows\footnote{The symbole $\asymp$ means logarithmic equivalence. Here, the limit $\varepsilon\to0$ is meant as $a_0\to 0$ at fixed $L$.}, 
\beqa \label{eq:large-dev}
\mathbb{P}\mathrm{rob}[\fn(\cdot)=n(\cdot)] \asymp_{\varepsilon\to0} e^{ - \varepsilon^{-1}\,F[n]} ~,
\eeqa
with $F[n]$ the so-called rate function. It plays a role analogue to the free energy but in non-equilibrium situations, and reduces to it at equilibrium. Out-of-equilibrium, $F[n]$ becomes a non-linear non-local functional of the density profile coding for long range correlations.

Equivalently, one can introduce the generating function of connected correlation functions of the density $\mathbb{E}\big[e^{\varepsilon^{-1}\! \int \! dx\,  h(x) \fn(x)} \big]$, with $h(x)$ some test function. By expanding in power of $h$, it generates the density correlation functions. The large deviation principle \eqref{eq:large-dev} implies that it scales as follows, 
\beqa \label{eq:Wh-dev}
\mathbb{E}_{\mathrm{ssep}}\big[ e^{ \varepsilon^{-1} \int \! dx\, h(x)\fn(x) } \big] \asymp_{\varepsilon\to 0} e^{\varepsilon^{-1} \,W[h] } ~,
\eeqa
for large system size. The two functions $F[n]$ and $W[h]$ are related by Legendre transform~: $W[h] = \mathrm{max}_{\{n(x)\}} \big[ \int_0^1 dx\, h(x)n(x) - F[n]\big]$. Formula \eqref{eq:Wh-dev} implies that the  cumulants of the density of order $P$ scale like $\varepsilon^{P-1}$ in the large system size limit.

For SSEP, the large deviation function $W[h]$ is given as the solution of an extremization problem~: 
\beqa \label{eq:Wh-int}
 W[h] =  \int_0^L \!\!\! dx\big[ \log\big(1+g(x)(e^{h(x)}-1)\big) - \log({g'(x)}L/{\Delta n})\big] ~,
 \eeqa
with $g(x)$ solution of the non-linear differential equation,
\beqa  \label{eq:Wh-extrem}
 \big(1+g(x)(e^{h(x)}-1)\big)g''(x) = g'(x)^2(e^{h(x)}-1) ~, 
 \eeqa
with boundary conditions $g(0)=n_a$ and $g(L)=n_b$. This condition is the Euler-Lagrange equation for $W[h]$ to be extremal with respect to variations of $g$. Expanding $W[h]$ in power of $h$ yields the first few density cumulants~: 
\beqa \label{eq:corr-ssep}
\mathbb{E}_\mathrm{ssep}[\fn(x)]= \bar n(x) ~,\quad
\mathbb{E}_\mathrm{ssep}[\fn(x)\fn(y)]^c= - \varepsilon\, (\Delta n)^2\, x(L-y)/L^2~,\quad \mathrm{for}\ x<y,
\eeqa
with $\bar n(x)$ the linear profile interpolating the two boundary densities, $\bar n(x)=(n_a (L-x) + x n_b)/L$.  In particular, the mean current $\bar \fj = -\partial_x\bar n = -\Delta n/L$, is proportional to the density gradient. Eq.\eqref{eq:corr-ssep} reveals the presence of long range, but decreasing with the system size, density correlations. 

The SSEP large deviation function \eqref{eq:Wh-int} was also shown to follow from the MFT \cite{Bertini2015MFT}. In its scaling limit, SSEP is mapped onto MFT with $D_\mathrm{ssep}=1$, $\sigma_\mathrm{ssep}(\fn)=\fn(1-\fn)$, and free energy $f_\mathrm{ssep}(\fn)=\fn\log\fn+(1-\fn)\log(1-\fn)$. The MFT has however a wider range of applications as it applies to all classical, locally diffusive, systems. It is worth stressing that MFT, formulated as an effective fluctuating hydrodynamics, depends only on two phenomenological coefficients~: the diffusion coefficient $D(\fn)$ and the mobility $\sigma(\fn)$, which are near equilibrium data.

\subsection{What a Quantum Mesoscopic Fluctuation Theory could be aiming at~?}

As the classical MFT, the {\it Quantum Mesoscopic Fluctuation Theory} (Q-MFT) should describe transports and their fluctuations in out-of-equilibrium systems. To be applicable to quantum systems, it should also address specifically quantum effects such as interference, coherence, entanglement, etc. 

The Q-MFT is expected to be applicable to locally diffusive but quantum systems. As a consequence, for the systems to be diffusive, not ballistic, Q-MFT should be valid at scales above the mean free path. For quantum effects not to be wash-out by decoherence, it should be applicable at scales below the decoherence length. This intermediate scale domain is called the mesoscopic domain.

At a sufficiently coarse-grained level, transports may usually be described by some hydrodynamic equations, even for genuinely quantum systems as in the Generalized Hydrodynamics (GHD) \cite{Castro2016emergent,Bertini2016Transport} adapted to integrable systems. Since they refer to locally conserved quantities, these equations take the form of local conservation laws, say of the form $\partial_t\fn + \partial_x \fj =0$, completed with constitutive equations expressing the current $\fj$ in terms of the density $\fn$ plus possible noise. This is also the way the classical MFT is formulated. Assuming decoherence, hydrodynamics hypothesises some kind of local equilibrium so that the system density matrices, reduced to the local hydrodynamic cells, are diagonal in the basis of the transported conserved charges. However, the off-diagonal elements survive at mesoscopic scales, even-though they are sub-leading compared to the diagonal ones. Usual hydrodynamics describes neither the structure of coherence or entanglement, nor their statistics or their dynamics. Describing them requires extracting information on the dynamics of the off-diagonal elements. Phrased differently, if Q-MFT could be formulated, it is expected to go beyond standard fluctuating hydrodynamics to include off-diagonal contributions.

At mesoscopic scales, quantum systems interact with extra noisy degrees of freedom leading to diffusion. For large systems, decoherence is hence at play so that the average dynamics is quasi-classical. Quantum interference effects echo fluctuating, non-classical, off-diagonal coherences. To make them manifest thus requires an understanding of coherent fluctuations, which cancel out on average and are sub-leading in the system size, and their noisy dynamics. For large systems, the statistical properties of these fluctuations can often be described by large deviation functions which code for rare fluctuations away from the typical value. Of particular interest are the large deviation functions say for quantum expectations of operators sensitive to quantum coherences or for entanglement entropies, which are specific to quantum systems, (see exemples of such large deviation functions in the closed Q-SSEP in Section \ref{sec:closed-details}).

In the last few years, impressive progresses on closely related topics has been achieved in critical or integrable quantum systems. A precise picture of entanglement and quenched dynamics has been obtained in terms of quasi-particle dynamics \cite{calabrese2016quantum}. A good understanding of transport and out-of-equilibrium phenomena in those systems has been achieved using the Generalized Hydrodynamics we alluded to above. GHD also lead to a good description of fluctuation statistics of transport in integrable systems \cite{Doyon2019Fluctuations} (see \cite{Bernard2012Energy} for similar studies in critical systems). However, these understandings are restricted to predominantly ballistic systems, even-though sub-leading diffusive effects have been incorporated \cite{DeNardis2018Hydrodynamic}. 

A naive way to look for a quantum version of MFT is to quantize equations \eqref{eq:mft}. Since eq.(\ref{eq:mft-b}) is a constraint, expressing the current $\fj$ in terms of the density $\fn$ plus noise, a direct quantization of eqs.(\ref{eq:mft}) is however potentially difficult, because, in a quantum theory, a constraint should be promoted to an operator identity. Following the strategy recently applied to GHD \cite{Ruggiero2020QGHD}, a possible route could amount to look for fluctuations above a classical solution $(\fn_\mathrm{cl}, \fj_\mathrm{cl})$ of the MFT equations \eqref{eq:mft}, say in form
\[ \fn =  \fn_\mathrm{cl} +\partial_x\varphi~,\quad \fj =  \fj_\mathrm{cl} -\partial_t\varphi ~, \]
to fulfil the continuity equation \eqref{eq:mft-a}, and to quantize the resulting Gaussian theory for the field $\varphi$. Another route could consist in promoting eq.\eqref{eq:mft-b} to a dynamical equation. For instance \cite{Bauer2017Stochastic}, eq.\eqref{eq:mft-b} could be upraised to a dynamical one by introducing a current friction, say 
\[ \tau_f\,\partial_t \fj = -D( \fn) \partial_x \fn - \eta\, \fj + \varepsilon^{1/2}\, \sigma(\fn)^{1/2}\, \xi ~, \]
with $\eta$ a dimensionless control parameter and $\tau_f$ a time scale parameter, so that the current friction coefficient is $\eta /\tau_f$. In the large friction limit, $\eta\, \fj \gg \tau_f \partial_t \fj$, this equation reduces to eq.\eqref{eq:mft-b}, but it can now be quantized.

However, both approaches fail to answer questions about the nature of the noise Q-MFT should incorporate. More importantly, there is also no a priori guarantee that such direct quantization of the MFT appropriately describes quantum coherent effects and their fluctuations in diffusive extended systems. Therefore, we will advocate below another route consisting in analysing simple microscopic models, say quantum variants of the classical exclusion processes, and in deducing, if possible, patterns for a Q-MFT applicable to a large class of mesoscopic diffusive systems.

\subsection{How do stochastic effects impact quantum systems~?}

The previous discussion points towards analysing noisy dynamics in Quantum Mechanics, in particular in extended systems. There are (at least) three main, interconnected, ways in which probabilistic phenomena hit Quantum Mechanics~:
\begin{itemize}

\item{} {\it Intrinsically}, as the results of measurements on quantum systems. Indeed, any information extracted by monitoring a quantum system is random, as expressed by the Born's rule of Quantum Mechanics, and it impacts the quantum system by the principle of quantum measurement back-action, even for weak measurements. As a consequence, the evolutions of monitored quantum systems are stochastic. They are actually described by the so-called quantum trajectories which are instrumental in controlling quantum systems, see e.g. \cite{wiseman_milburn_2009,Jacobs2014Book}.

\item{} {\it Extrinsically}, as noise, providing stochastic models for environments interacting with quantum systems. Any system interacting with an environment is subject to a noisy evolution. Noise, classical or quantum, may be viewed as the collection of the degrees of freedom of the -- not precisely known and controlled -- environments. As a consequence, random dynamics arise in Quantum Mechanics as an efficient way of describing the evolution of systems interacting with environments or external fields, as in the original model of bosonic baths introduced by Caldeira and Leggett \cite{Caldeira1981Influence}.

\item{} {\it Generically}, as models for typical, alias generic, quantum states and operations. This is for instance the principle underlying the use of random matrix theory (RMT) in Quantum Mechanics or Quantum Information, as first advocated by Wigner, see e.g. \cite{Mehta2004RMT}. More recently, this is also the way random quantum circuits have been used to study universal behaviours of quantum chaotic systems.   By adding stochasticity, these systems ought to lose their fine properties pertaining to particularities, such as specific symmetries, thus allowing the emergence of generic properties.
\end{itemize}

Analysing quantum simple exclusion processes fits in the two last categories. The latter are models of stochastic many-body dynamics incorporating diffusive effects and their fluctuations. Although one may imagine experimental realisations\footnote{One may look for spin chains subjected to strong noisy magnetic fields, since Q-SSEP describe the effect dynamics of the latter in the limit of large magnetic fields. }, the purpose of these model systems are, as any toy model, to propose paradigmatic models, which for instance allow exact results to be derived, and to reveal patterns which are potentially generic for mesoscopic diffusive quantum systems. Maybe optimistically, the conclusions then obtained lead to a simple framework, depending on only a few number of characteristics and parameters, as the MFT does, and applicable to a large enough universality class of physical systems.

Considering the impact of monitoring on these systems is certainly important as it has been recently done in quantum circuit models. We will however not discuss them in this short review.

\subsection{How can stochastic fluctuations be observed~?}

When observing or measuring a quantum system, one gets random outputs whose distribution is encoded in the system density matrix $\rho$ via Born's rule~: the probability for an output $o$ is $p_o= \Tr(\rho\, \Pi_o)$, if the measured observable is $O$ and $\Pi_o$ its spectral projector on the eigenvalue $o$. A way to measure these distributions is to prepare a large number $\mathcal{N}$ of copies of the system, say $\mathcal{N}\approx 10^3$, $10^6$ or bigger, and to repeat the measurement experiment under identical conditions. The output histogram is then obtained by collecting the measurement outputs $o_1,o_2,\cdots, o_{\mathcal{N}}$ and counting the number of times $\mathfrak{n}_o$ a given output $o$ appears in this list. Then $\mathfrak{n}_o/\mathcal{N}\simeq p_o$ for $\mathcal{N}\gg1$. Given these data, one has access to the expectation of the observable $O$, or its higher moments, via $\vev{O^k}=\sum_o o^k\mathfrak{n}_o/\mathcal{N} = \Tr(\rho\, O^k)$. This is nicely illustrated in Haroche's photon box experiment~\cite{Guerlin2007ProgressiveFC}.

Suppose now that the system evolves in time and that its preparation or its evolution is noisy, so that the system density matrix is time dependent and random. To make it explicit (when needed), we shall write $\rho_t(\omega)$ with $\omega$ some variables representing possible events and taking values in some probability space. When iterating measurement processes, two different setups can be envisioned to get an estimation of the output histograms for an observable $O$~:\\
-- (i) either one has some control on the noise, so that one can prepare copies of the systems and let them evolve under identical noisy conditions. Then the outputs $o_1,o_2,\cdots, o_{\mathcal{N}}$ are distributed in a way depending on the noise sample, say $\omega$, so that their histogram is
\beqa
p_o(\omega)=\mathfrak{n}_o(\omega)/\mathcal{N} \simeq \Tr(\rho_t(\omega)\, \Pi_o)~,\quad \mathrm{for}\ \mathcal{N}\gg1 ~,
\eeqa
with $\mathfrak{n}_o(\omega)$ the numbers of time the value $o$ appear in the output lists for all copies of the system\footnote{This may alternatively be formulated in terms of conditioned probability as $p_o(\omega)$ is the probablity to observe the value $o$ conditioned on the noise being in the state $\omega$.} under identical noisy conditions $\omega$. \\
-- (ii) or one does not control the noise so that the latter is sampled differently at each iteration of the measurement process. Then, if the number of iterations $\mathcal{N}$ is large enough, and much larger than in the previous setup, the noisy variables are sampled faithfully according to some probability measure $p_\omega$ so that the output histogram is 
\beqa
\bar p_o =\bar{\mathfrak{n}}_o/\mathcal{N} \simeq \Tr(\bar \rho_t\, \Pi_o)~,\quad \bar \rho_t := \sum_\omega p_\omega \rho_t(\omega) ~,\quad \mathcal{N}\gg1 ~.
\eeqa

Phrased differently, by not controlling the noise at the different iterations of the measurement process, we only have access to the average density matrix, $\bar \rho_t= \sum_\omega p_\omega \rho_t(\omega)$, and to the average quantum expectation values~: 
\beqa
 \mathbb{E}[\vev{O}_t] := \Tr(\bar \rho_t\, O) = \sum_\omega p_\omega\, \Tr(\bar \rho_t(\omega)\, O) ~.
 \eeqa
 
Under noisy conditions, the histograms $p_{o}(\omega)=\mathfrak{n}_o(\omega)/\mathcal{N}$ may fluctuate from one noise realisation to the other. A way to quantify the fluctuations of the quantum expectation values is to evaluate their correlations, say, 
\beqa \label{eq:noisy-fluct}
 \mathbb{E}[\vev{O^{(1)}}_t\vev{O^{(2)}}_t] := \sum_\omega p_\omega\, \Tr(\rho_t(\omega)\, O^{(1)})\Tr(\rho_t(\omega)\, O^{(2)}) ~,
 \eeqa
for two observables $O^{(1)}$ and $O^{(2)}$. This requires producing the two histograms $p^{(1)}_{o_1}(\omega)$ and $p^{(2)}_{o_2}(\omega)$ by iterating measurements of the two observables $O^{(1)}$ and $O^{(2)}$ under identical noisy conditions\footnote{Actually, this assertion is also true in classical physics where to estimate the correlations between two quantities $X$ and $Y$ one has to sample their possible values $x(\omega)$ and $y(\omega)$ under identical noisy conditions in order to estimate their correlations $\mathbb{E}[XY]=\sum_\omega p_\omega x(\omega) y(\omega)$. However, the consequences of these samplings in quantum mechanics is more dramatic due to the back-action of the measurement process on the system.}.

Although controlling the noise may be difficult experimentally and even looks paradoxical -- since, more or less by definition, noise is a name for variables whose behaviors cannot be controlled -- this is not completely unimaginable. For instance, one may imagine placing copies of the system under the umbrella of the same noise, say if the randomness arises from an external random magnetic field\footnote{A field frequency spectrum containing a few incommensurable frequencies would be generate a field whose time evolution is closed enough to a good realisation of a white-noise.}. Or, if the noise is artificially produced, say by random number generators, one may simply ensure that the generators are implemented using identical seeds. Or, if the fluctuations arise from variances in the initial state, one may control this initial state, etc. 

Nevertheless, as for classical systems, having a realistic, generic, controlled models for the noise and the fluctuations it induces, allows to estimate the fluctuations due to noisy environments that are not controlled. 

Taking the time evolution of the system into account, one may wonder how these fluctuations evolve in time and whether they survive the large time limit and/or the large size limit, especially for ergodic systems. A commonly used formulation of ergodicity, especially in the physics literature on many-body physics, is to assume that the time averaged system density matrix converges at large time to some steady density matrix $\rho_\infty$ (supposed to be unique for simplicity)\footnote{This notion of ergodicity is not that used in the mathematical literature, say on quantum billards, which is  grounded on semi-classical structures.}:
\beqa \label{eq:ergodicity1}
\lim_{T\to\infty} \frac{1}{T} \int_0^T \hskip -2truemm ds\, \Tr(\rho_s\, O) = \Tr(\rho_\infty\, O) ~.
\eeqa
for some appropriate set of observables $O$. If the system is noisy, one hypothesises self averaging and extend the above definition to an almost sure convergence. Defining ergodicity via eq.\eqref{eq:ergodicity1} leads to wonder (say, because of the measurement back-action) how the time integral $\int_0^T \! ds\, \Tr(\rho_s\, O)$ may be estimated experimentally. However, in the context of monitored systems and their quantum trajectories  \cite{wiseman_milburn_2009,Jacobs2014Book}, such almost sure convergence is known to hold under quite general hypothesis~\cite{Kuemmerer2004APE,Kuemmerer2001AnET}.

At a classical level, if $x\to x_t$ is a dynamical flow on some space $\mathfrak{X}$, ergodocity w.r.t. to some measure $d\nu_\infty$ on $\mathfrak{X}$ is the property that $\lim_{T\to\infty} T^{-1}\int_0^T\hskip -1truemm ds\, f(x_s)=\int_\mathfrak{X} d\nu_\infty(x)\, f(x)$, for appropriate functions $f$. At the quantum level, there are two options :
(i) either one views quantum observables as quantization of functions on some phase space and density matrices as quantization of measures on that space, then eq.\eqref{eq:ergodicity1} is a natural analogue of the classical ergodicity (as long as measurements are not taken into account, see Appendix \ref{app:conjecture});
(ii) or one considers the density matrices $\rho$ as the dynamical variables, then one would like to apply ergodicity to non-linear functions of $\rho$.

Considering density matrices as the dynamical variable, eq.\eqref{eq:ergodicity1} should better be called {\it linear ergodicity} as it only tests linear functions of the system density matrix. More generally, to have a hand on fluctuations, one may consider more general functions and ask that the time averages of say all polynomials in the density matrix (up to a given degree) converge, that is~:
\beqa \label{eq:ergodicity2a}
\lim_{T\to\infty} \frac{1}{T} \int_0^T  \hskip -2truemm ds\, \Tr(\rho_s\, O^{(1)})\cdots\Tr(\rho_s\, O^{(p)}) ~\quad \mathrm{exist}.
\eeqa 
{\it Non-linear ergodicity} would hold if there exists a steady measure $\mathbb{E}_\infty$ on density matrices such that these large time limits are the moments of the density matrix w.r.t. this measure, that is:
\beqa \label{eq:ergodicity2bis}
\lim_{T\to\infty} \frac{1}{T} \int_0^T  \hskip -2truemm ds\, \Tr(\rho_s\, O^{(1)})\cdots\Tr(\rho_s\, O^{(p)}) = \mathbb{E}_\infty[\Tr(\rho\, O^{(1)})\cdots\Tr(\rho\, O^{(p)})]~,
\eeqa 
If the system evolution is noisy, one may require that this property holds almost surely. Of course, we can also extend this definition by considering multi-time expectations. 

Checking ergodicity experimentally, even at the linear level as in eq.\eqref{eq:ergodicity1}, by producing the appropriate measurement histograms requires double series of copies $\mathcal{M}\times\mathcal{N}$ of the system: $\mathcal{N}$ copies to produce the measurement histograms at each given instant, repeated $\mathcal{M}$ times to sample the time interval  from $0$ to $T$. The difficulties are comparable to those of the direct approach to evaluate the noisy fluctuations \eqref{eq:noisy-fluct}. The situation is better if one imagines monitoring the quantum system. See the Appendix \ref{app:conjecture}.

For extended systems, assuming that a steady regime is attained at large time, we expect the measure $\mathbb{E}_\infty$ to be peaked around a mean density matrix $\bar \rho_\infty$, with fluctuations $\delta\rho_t$ decreasing with the system size  ~:
\beqa \label{eq:rho-infty}
\rho_t \simeq \bar \rho_\infty + \delta\rho_t ~,\quad \mathrm{at\ large\ enough\ time}, 
\eeqa
or more precisely, $\Tr(\rho_t\, O)\simeq Tr(\bar \rho_\infty\, O) + Tr(\delta\rho_t\, O)$, for appropriate observables $O$. These fluctuations are expected to be small and to scale as $1/L^\alpha$, with $L$ the macroscopic linear size of the system and $\alpha>0$. The law of large number suggests $\alpha=\frac{1}{2}$. As a consequence, we expect that time averaging is unnecessary so that the density matrix, tested appropriately, converges at large time,
\beqa \label{eq:converge-1}
\lim_{t\to\infty} \Tr(\rho_t\, O) = \Tr(\bar \rho_\infty\, O) ~,\quad \mathrm{in\ infinite\ volume}, 
\eeqa
for an appropriate set of observables, say local observables. This is a stronger convergence than that of linear ergodicity \eqref{eq:ergodicity1}. This is the expected behaviour when thermalisation holds but also in some non-equilibrium situations.

\begin{figure}\bc
\includegraphics[width=0.45\textwidth]{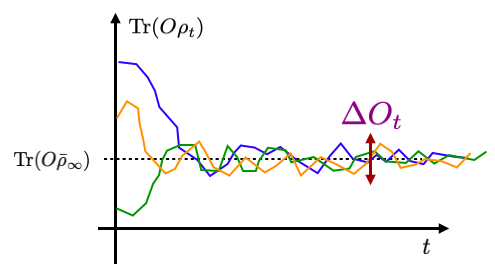}\ec
\caption{\it Schematic representation of the convergence of quantum expectation values to their equilibrium values with sub-leading fluctuations $\Delta O_t:=\Tr(O\delta\rho_t)$ decreasing with the system size.}
\label{fig:fluctu-time}
\end{figure}

There are fluctuations, in time and from sample to sample, when approaching the asymptotic state $\bar \rho_\infty$. See Figure \ref{fig:fluctu-time}. At large time, but at finite (large) volume, they are expected to reach a steady distribution coding for correlations between quantum expectation values (at large time), say 
\[ 
\mathbb{E}_\infty[\Tr(\rho\, O^{(1)})\cdots\Tr(\rho\, O^{(r)})] ~.
\] 
Deviation from $ \Tr(\bar \rho_\infty\, O^{(1)})\cdots\Tr(\bar \rho_\infty\, O^{(r)})$ are sub-leading in the system size. At finite but large enough volume, we may expect that their evolution acquires a universal status, so that they can be described by some effective, universal, noisy dynamics. Having good noisy dynamical models then provides a way to have a hand on those fluctuations and their time evolutions.

A natural question is then how could we estimate these fluctuations by looking at a single time trajectory of the system state. A way to proceed could consist in slicing the time axis into pieces of time duration of order $\tau$, with $\tau$ an effective correlation time such that the average over time scale of order $\tau$ smoothes out the fluctuations, i.e.  $\bar \rho_t:= \frac{1}{\tau} \int_{t}^{t+\tau} ds\, \rho_s$ has a smooth time evolution. This smoothed mean state $\bar \rho_t$ at time $t$ can be evaluated by sampling the time interval $[t,t+\tau]$ at a large number of times $t_1,\, t_2,\cdots, t_\mathcal{P}$, that is: $\Tr(\bar \rho_t\, O)=\frac{1}{\mathcal{P}}\sum_{j=1}^\mathcal{P} \Tr(\rho_{t_j}\, O)$. This sampling yields to an estimation of the fluctuations at time $t$ via
\[
\frac{1}{\mathcal{P}}\sum_{j=1}^\mathcal{P} \Tr(\rho_{t_j}\, O^{(1)})\cdots\Tr(\rho_{t_j}\, O^{(r)}) ~,
\]
with all  $t_j\in[t,t+\tau]$ and $\mathcal{P}$ large enough.
We could then declare that the above mentioned effective noisy dynamics is faithful if its moments coincide with these correlations, that is~: 
\[ 
\mathbb{E}[\Tr(\rho\, O^{(1)}_t)\cdots\Tr(\rho\, O^{(r)}_t)] \stackrel{?}{=}
\frac{1}{\mathcal{P}}\sum_{j=1}^\mathcal{P} \Tr(\rho_{t_j}\, O^{(1)})\cdots\Tr(\rho_{t_j}\, O^{(r)})~,
\]
At the linear level, for $t$ large enough, this reproduces the convergence condition \eqref{eq:converge-1} with $\bar \rho_\infty$ the mean steady state of the effective dynamics. 

\section{Quantum Simple Exclusion Processes}

The quantum simple exclusion processes are models of quantum many-body noisy dynamics describing fermions hopping stochastically along a one dimensional chain\footnote{Quantum simple exclusion processes can be defined on any graph.}. The chain can be closed with periodic boundary conditions, or open with injection and extraction of particles at its two ends. The processes can be symmetric (resp. asymmetric) depending whether the amplitudes to move to the right or to the left are equal (resp. different). The exclusion constraint is implemented by the fermionic character of the degrees of freedom. They are quantum extensions of the classical simple exclusion processes.

\subsection{Q-SSEP and Q-ASEP}

\begin{figure}\bc
\includegraphics[width=0.65\textwidth]{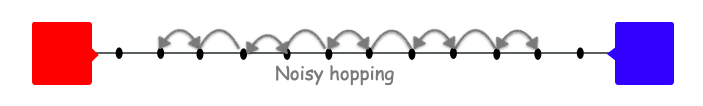}\ec
\caption{\it Schematic representation of Q-SSEP or Q-ASEP. Periodic boundary conditions are imposed in the closed case (no reservoirs). In the open case, injection/extraction processes at the two ends represent contact with two reservoirs, here depicted in red or blue.}
\label{fig:qssep-setup}
\end{figure}

\hskip 0.6 truecm $\bullet$ {\it Closed Q-SSEP.}\\
In the closed symmetric setup, the system dynamics is unitary but noisy. The system density matrix $\rho_t$ evolves unitarily, $\rho_{t+dt} = e^{-idH_t}\rho_t\, e^{idH_t}$,
with $dH_t$ the hamiltonian increment between time $t$ and $t+dt$. For Q-SSEP, the hamiltonian increment is defined as\footnote{This definition can be generalized by including $cc$ or $c^\dag c^\dag$ terms not preserving the particle number but still quadratic. However, transport is associated to locally conserved quantities and such models might not be well adapted to study transport.}
\beq \label{eq:def-dH}
 dH_t := \sqrt{J}\, \sum_{j=1}^{N} \big( c^\dag_{j+1}c_{j}\, dW_t^j + c^\dag_{j}c_{j+1}\, d\overline{W}_t^j \big) ~,
 \eeq
 where $c_{j}$ and $c_{j}^{\dagger}$ are canonical fermionic operators, one pair for each site of the chain, with $\{c_{j},c_{k}^{\dagger}\}=\delta_{j;k}$, and $W_{t}^{j}$ and  $\overline W_{t}^{j}$ are pairs of complex conjugated Brownian motions, one pair for each edge along the chain, with zero mean, $\mathbb{E}[dW^j_t]=\mathbb{E}[d\overline{W}^j_t]=0$, and covariance $\mathbb{E}[dW_{t}^{j}d\overline{W}_{t}^{k}]=\delta^{j;k}\,dt$.
Here, $N$ is the number of sites on the chain and  $J$ a bare coupling constant, with the dimension of a frequency. For the closed Q-SSEP,  periodic boundary conditions are assumed. See Figure \ref{fig:qssep-setup}.

Upon discretizing the time evolution, this model can be viewed as a random quantum circuit with Gaussian two-site gates. In this discrete setting, after $n$ time steps, each of duration $\delta t$, the system density matrix $\rho_n$  is unitary up-dated according to $\rho_n \to \rho_{n+1} = U_n\rho_n U_n^\dag$ with $U_n=V^1_nV^0_n$ where the $V_n^e$'s (with $e=0,1$ for even and odd) are products of gates acting on adjacent sites (for $N=2M$)~:
\beqa
V_n^e &=& W^{(n)}_{e+1,e+2}W^{(n)}_{e+3,e+4}\cdots W^{(n)}_{e+N-1,e+N} ~,\\
W^{(n)}_{j,j+1} &=& \exp\big({c^\dag_{j+1}c_{j}\, \xi^{(n)}_{j} + c^\dag_{j}c_{j+1}\, \bar\xi^{(n)}_{j} }\big) ~, \nonumber
\eeqa
with $\xi^{(n)}_{j}$ independent identically distributed (i.i.d.) Gaussian variables with zero mean and variance $\mathbb{E}[\xi^{(n)}_{j}\bar \xi^{(m)}_{k}] =\delta^{n;m}\delta_{j;k}\, J\delta t$. This discretization yields to simple numerical implementations of the dynamics. In the following we shall however stick to the time continuous formulation (which is better adapted to analytical descriptions).

Since $\rho_{t+dt} = e^{-idH_t}\rho_t\, e^{idH_t}$, the equation of motion for the system density matrix reads (with It\^o convention)~:
\beq \label{eq:motion-rho}
d\rho_t = -i[dH_t,\rho_t] - \frac{1}{2} [ dH_t,[dH_t,\rho_t]] ~.
\eeq
These equations are classical stochastic differential equations (SDE) but on quantum density matrices. The double commutator in eq.\eqref{eq:motion-rho} is an echo of the fact that the Brownian increments $dW_{t}^{j}$ scale as $\sqrt{dt}$. The equation of motion for the observables, say $O$, are defined from eq.\eqref{eq:motion-rho} by duality: $\Tr(\rho_t\,O)= \Tr(\rho\, O_t)$. In particular, the equation of motion for the local particle number operators, $\hat n_j:=c_j^\dag c_j$, reads
\beq \label{eq:n-diff}
 d\hat n_j = J\, \Delta\hat n_j\, dt + \sqrt{J}\big(d\hat{\mathbb{V}}_s^j-d\hat{\mathbb{V}}_s^{j-1}\big) ~,
\eeq
with $\Delta$ the discrete Laplacian, $\Delta \hat n_j := \hat n_{j+1}-2\hat n_j+\hat n_{j-1}$, and $d\hat{\mathbb{V}}_t^j:=i\big( c_j^\dag c_{j+1}dW^j- \mathrm{h.c.}\big)$ noisy operators. Eq.\eqref{eq:n-diff} indeed codes for stochastic diffusion for quantum observables. The Q-SSEP is one of the simplest model of pure, quantum, noisy, diffusive dynamics.

Since information is lost when averaging over the realizations of the noise, the average dynamics is not unitary but dissipative. In particular, the average number operator $\mathbb{E}[\hat n_j]$ satisfy to pure diffusion: $\partial_t\mathbb{E}[\hat n_j] = J\, \mathbb{E}[\Delta\hat n_j]$. By the Markov property of the Brownian motions, the average dynamics defines a semi-group generated by a Lindbladian which is obtained by averaging eq.\eqref{eq:motion-rho}~:
\beq \label{eq:L1-dyn}
\partial_t \bar \rho_t = \mathcal{L}_{\mathrm{ssep}}(\bar \rho_t) ~,
\eeq
with $\bar\rho_t:=\mathbb{E}[\rho_t]$ the average density matrix. The Lindbladian $\mathcal{L}_{\mathrm{ssep}}$ is the sum of local Lindbladians,  one for each edge of the chain, $\mathcal{L}_{\mathrm{ssep}}=J\, \sum_{j=1}^N {\mathcal{L}}_{j;j+1}$. Each of them decomposes as $\mathcal{L}_{j;j+1}=\overrightarrow{\mathcal{L}}_{j;j+1} +  \overleftarrow{\mathcal{L}}_{j;j+1}$, where 
\begin{subequations}\label{eq:L-edge}
\begin{align} 
\overrightarrow{\mathcal{L}}_{j;j+1}(\rho)  &:= \ell^-_{j}\, \rho\, \ell^+_{j} -\frac{1}{2}(\ell^+_{j}\ell^-_{j}\, \rho + \rho\,\ell^+_{j}\ell^-_{j}) ~,\\
\overleftarrow{\mathcal{L}}_{j;j+1}(\rho)  &:= \ell^+_{j}\, \rho\, \ell^-_{j} -\frac{1}{2}(\ell^-_{j}\ell^+_{j}\, \rho + \rho\,\ell^-_{j}\ell^+_{j}) ~,
\end{align}
\end{subequations}
with $\ell^+_{j} := c^\dag_{j+1}c_{j}$ and $\ell^-_{j} :=c^\dag_{j}c_{j+1}$. The local Lindbladians $\overrightarrow{\mathcal{L}}_{j;j+1}$ (resp. $\overleftarrow{\mathcal{L}}_{j;j+1}$) describe jumps of the fermions to the right (resp. left).
Higher moments of the density matrix, say its quadratic fluctuations $\mathbb{E}[\rho_t\otimes\rho_t]$, also satisfy Lindblad dynamics~\cite{Bernard2021Dynamics}. One of the motivation for studying Q-SSEP is to get a good description of quantum diffusive dynamics, beyond the average dynamics, including fluctuations. See below section \ref{sec:closed-fluctuation}.

The name Q-SSEP is inherited from the fact that the mean dynamics, when reduced to an appropriate sector, is that of the classical SSEP. At each site along the chain, the empty $\ket{\bullet}$ and full $\ket{\oo}$ states, with respectively one and zero fermion, form a basis of states and diagonalize the particule number operator (with eigen-value $1$ or $0$). The states $\ket{\bm{n}}$ diagonalising all the particle numbers along the chain are thus indexed by the classical configurations $\bm{n}=(\nu_1,\cdots,\nu_N)$, with $\nu_j=0,1,$ the particle number at site $j$.
 A density matrix diagonal in this particle number basis specifies a probability measure on classical configurations since it can be written as $\rho_\mathrm{diag} = \sum_{\bm{n}} Q_{\bm{n}}\, \Pi_{\bm{n}}$, with $\Pi_{\bm{n}}:=\ket{\bm{n}}\bra{\bm{n}}$ and $Q_{\bm{n}}$ a probability measure on $\bm{n}$~: $\sum_{\bm{n}} Q_{\bm{n}}=1$, $Q_{\bm{n}}\geq0$. The mean Q-SSEP dynamics \eqref{eq:L1-dyn} preserves diagonal density matrices and thus defines a flow -- a Markov chain -- on probability measures on classical configurations. This flow is identical to that of the classical SSEP since
\begin{align*}
\mathcal{L}_{\mathrm{ssep}} (\ket{\oo\oo}\bra{\oo\oo}) & =0 ~,\\
\mathcal{L}_{\mathrm{ssep}}(\ket{\oo\bullet}\bra{\oo\!\bullet\!}) & = J( -\ket{\oo\bullet}\bra{\oo\!\bullet\!}+\ket{\!\bullet\!\oo}\bra{\!\bullet\oo} )~,\\
\mathcal{L}_{\mathrm{ssep}}(\ket{\!\bullet\!\oo}\bra{\bullet\oo}) & = J( +\ket{\oo\bullet}\bra{\oo\!\bullet\!}-\ket{\!\bullet\!\oo}\bra{\!\bullet\oo} )~,\\
\mathcal{L}_{\mathrm{ssep}}(\ket{\!\bullet\!\bullet}\bra{\!\bullet\!\bullet\!}) & =0 ~.
\end{align*}
This can be compared with the Markov transition matrix of classical SSEP in eqs.\eqref{eq:P-ssep-classic}. 
As a consequence, the generating function of the steady fluctuations of the classical SSEP occupancies $\mathfrak{n}_j$ can be expressed as a quantum expectation value w.r.t. the steady average Q-SSEP density matrix~:
\beqs \label{eq:class-SSEP}
\mathbb{E}_\mathrm{ssep}[ e^{\sum_j h_j \mathfrak{n}_j} ] = \mathbb{E}\Big[ \mathrm{Tr}\big( \rho\, e^{\sum_i h_i \hat n_i}\big)\Big]= \mathrm{Tr}\big( \bar \rho\, e^{\sum_i h_i \hat n_i}\big) ~,
\eeqs
with $\hat n_i=c^\dag_ic_i$ the quantum number operators, $\bar \rho=\mathbb{E}[\rho]$ the mean Q-SSEP state, averaged w.r.t. the Q-SSEP steady measure.

The quantum Q-SSEP is however an extension of the classical SSEP as this correspondance only applies to the average dynamics of diagonal density matrices. For instance, while only classical configurations are considered in classical SSEP, cats like states made of superpositions of classical configurations are allowed in the Q-SSEP, say
\[
\ket{\cdots \oo\oo\!\bullet\!\bullet\oo\!\bullet\!\oo\oo\!\bullet\cdots} + \ket{\cdots \bullet\!\oo\!\bullet\!\oo\oo\!\bullet\!\oo\!\bullet\!\bullet\cdots} + \cdots ~,
\]
even-though the states describe identical density profiles at a coarse-grained level. These states could interfere and Q-SSEP describes their statistics and dynamics.

\medskip

$\bullet$ {\it Open Q-SSEP.}\\
The open Q-SSEP is defined similarly except that the chain interval is now open (no periodic boundary conditions) and the dynamics is modified by including injection and extraction processes at the two ends of the chain. These processes are represented by dissipative, deterministic, Lindblad dynamics. As a consequence, the equations of motion for the system density matrix read~:
\beq \label{eq:Q-flow}
 d\rho_{t}=-i[dH_{t},\rho_{t}] -\frac{1}{2}[dH_{t},[dH_{t},\rho_{t}]]+\mathcal{L}_\mathrm{bdry}(\rho_{t})dt ~,
 \eeq
with $dH_t$ as above in eq.\eqref{eq:def-dH}, and $\mathcal{L}_\mathrm{bdry}$ a boundary Lindbladian acting on sites $1$ and $N$, with $\mathcal{L}_\mathrm{bdry}=\alpha_{1}{\cal L}_{1}^{+}+\beta_{1}{\cal L}_{1}^{-}+\alpha_{N}{\cal L}_{N}^{+}+\beta_{N}{\cal L}_{N}^{-}$ and 
\begin{subequations}
\begin{align} \label{eq:L-bdry}
{\cal L}_{j}^{+}(\star) & =c_{j}^{\dagger}\star c_{j}-\frac{1}{2}(c_{j}c_{j}^{+}\star+\star c_{j}c_{j}^{\dagger}) ~,\\
{\cal L}_{j}^{-}(\star) & =c_{j}\star c_{j}^{\dagger}-\frac{1}{2}(c_{j}^{\dagger}c_{j}\star+\star c_{j}^{\dagger}c_{j}) ~,
\end{align}
\end{subequations}
where the parameters $\alpha_j$ (resp. $\beta_j$) are the injection (resp. extraction) rates.
These boundary processes drive the system out-of-equilibrium by enforcing a non-trivial current through the system. 

General comments about Q-SSEP apply to the open case, if the boundary terms are taken into account appropriately. In particular, the average dynamics is also of Lindblad type, with contributions from the boundary Linbladians. On diagonal density matrices, it reduces to that of the open SSEP, including the boundary terms. Higher moments also follow a Lindblad dynamics \cite{Bernard2021Dynamics}. Despite the presence of boundary dynamics, the open Q-SSEP preserves gaussianity of the states and it is thus particularly adapted to exact studies. 
Duality property of such model systems were pointed out in \cite{frassek2020duality}.

\medskip

$\bullet$ {\it Closed and open Q-ASEP.}\\
The Q-ASEP is slightly different from the Q-SSEP because the noise the chain is coupled to is not classical but quantum. The hamiltonian increments $dH_t$ are of the same form as in eq.\eqref{eq:def-dH}, namely,
\beq \label{eq:def-dH-bis}
 dH_t := \sqrt{J}\, \sum_{j=1}^{N} \big( c^\dag_{j+1}c_{j}\, dW_t^j + c^\dag_{j}c_{j+1}\, d\overline{W}_t^j \big) ~,
 \eeq
but where $W_t^j$ and $\overline{W}_t^j$ are now quantum Gaussian processes with zero mean and covariance,
\beq \label{eq:E-noise}
\mathbb{E}[dW^j_t\, d\overline{W}^k_t]= p\,\delta^{j;k}\,dt ~,\quad 
\mathbb{E}[d\overline{W}^j_t\, dW^k_t]= q\,\delta^{j;k}\,dt ~,
\eeq
with $p,\, q$, ($p\geq q$), two real positive parameters. The Q-ASEP correspond to $p\not= q$, and the Q-TASEP to either $p=0$ or $q=0$.  The two  processes $W^j_t$ and $\overline{W}^k_t$ are then non-commuting quantum operators, 
\beq
[dW^j_t,d\overline{W}^k_t]=(p-q)\,\delta^{j;k}\,dt ~.
\eeq

There are two physical ways to construct this noise. Either one may view them as the scaling limit of series of two-state probes, say Q-bit or spin half probes, recursively interacting with the system \cite{Attal2003FromRT}, in a way similar to the photon box experiment \cite{Guerlin2007ProgressiveFC}. This construction is parallel to that of the classical Brownian motion as a scaling limit of random walks.
Or, these noise can be directly constructed in the continuum, as in the theory of quantum stochastic differential equations \cite{Hudson1984Quantum}. In physical terms, the noise Hilbert space is the tensor product of noise Hilbert spaces attached to each edge of the chain, $\mathcal{H}_{\mathrm{noise}}=\otimes_j\mathcal{H}^{(j)}_{\mathrm{noise}}$. Each of them is a Fock space, over the space of square integrable functions on the line, on which complex bosonic fields, $\varphi^j(s)$ and  $\bar \varphi^j(s)$, are acting, with canonical commutation relations,
\beq
\big[ \varphi^j(s) , \bar \varphi^{k}(s') \big] = \delta^{j;k}\delta(s-s') ~.
\eeq
The noise expectation values are defined as quantum expectation values w.r.t. a particular state $\Omega_{\mathrm{noise}}$ on the noise Hilbert space. Averaging over the noise amounts to trace over the noise degrees of freedom, $\mathbb{E}[(\cdots)]:=\Tr(\Omega_{\mathrm{noise}}(\cdots)]$. The noise is chosen to be thermal so that $\mathbb{E}[\bar \varphi^j(s)\varphi^k(s')]=\mathsf{n}\,\delta^{j;k}\delta(s-s')$ with $\mathsf{n}$ some filling factor.
The quantum Brownian motions are then defined by
\beq
W^j_t=\lambda\!\int_0^t\! \! ds\, \varphi^j(s) ~,\quad \overline{W}^j_t=\lambda\!\int_0^t\! \!ds\, \bar \varphi^{j}(s) ~.
\eeq
Their increments are $dW^j_t=W^j_{t+dt}-W^j_t$. They satisfy eqs.\eqref{eq:E-noise} with $q=\lambda^2 \mathsf{n}$ and $p=\lambda^2(1+\mathsf{n})$. Noise expectation values are computed using Wick's theorem. The Brownian increments $dW^j_t$ and the portion of the noise Fock space associated to the fields $\varphi^j(s)$ between time $t$ and $t+dt$ represent the scaling limit of all probes which have interacted with the system during the time interval $[t,t+dt)$.

The noisy dynamics of observables, say $O$, generated by the hamiltonian increments \eqref{eq:def-dH-bis} are
\beq
dO_t = +i[dH_t, O]_t - \frac{1}{2}[dH_t,[dH_t,O]]_t ~,
\eeq
up to boundary terms in the case of an open chain. These are quantum stochastic differential equations as defined in \cite{Hudson1984Quantum}. For $O$ an observable on the system, its average w.r.t. the noise, $\bar O_t:=\mathbb{E}[O_t]$, satisfy a Lindblad equation
\beq
\partial_t\bar O_t= \mathcal{L}_{\mathrm{asep}}^\dag(\bar O)_t ~,
\eeq
with dual Lindbladian $\mathcal{L}_{\mathrm{asep}}^\dag=J\, \sum_{j=1}^N \big( p\overrightarrow{\mathcal{L}}^{\dag}_{j;j+1} +  q\overleftarrow{\mathcal{L}}^{\dag}_{j;j+1}\big)$, with $\overrightarrow{\mathcal{L}}^{\dag}_{j;j+1}$ and $\overleftarrow{\mathcal{L}}^{\dag}_{j;j+1}$ the duals of the Lindbladians \eqref{eq:L-edge}. This average dynamics was actually considered in \cite{Temme2012Stochastic,eisler2011crossover}. Higher moments also satisfy a Lindblad dynamics \cite{Bernard2021Dynamics}.

Since the two Lindbladians $\overrightarrow{\mathcal{L}}_{j;j+1}$ and $\overleftarrow{\mathcal{L}}_{j;j+1}$ code respectively  for left and right moves, the total Lindbladian $\mathcal{L}_{\mathrm{asep}}$ codes for asymmetric moves as in the classical ASEP. The exclusion principle is taken into account by the fermionic character of the particles. Restricted to diagonal density matrices, the average of Q-ASEP is that of the classical ASEP \cite{Jin2020Stochastic}. Q-ASEP is however an extension of the classical ASEP as it describes off-diagonal fluctuations and coherences.

In the following we restrict our discussion to the closed and open Q-SSEP.

\subsection{Closed Q-SSEP~: Fluctuations at equilibrium}
\label{sec:closed-fluctuation}

Q-SSEP is particularly adapted to exact analytic results because it is a quadratic model\footnote{But it is a non-trivial model because of the presence of noise.}, for each realization of the noise. In particular it preserves gaussianity of the system density matrix, again for each realization of the noise (not in average). The dynamics and statistics of such density matrices are encoded into that of their matrices of two-point functions $G_{ij}:=\Tr(\rho_t c_j^\dag c_i)$. Hence, analysing Q-SSEP reduces to studying the statistics and dynamics of its two-point functions. This reduces the number of degrees of freedoms from $2^N$  to $N^2$.
For the closed Q-SSEP, the time evolution of the matrix $G$ is also unitary,
\beq \label{eq:G-time}
G_{t+dt} = e^{-idh_t}\ G_{t}\, e^{idh_t} ~,
\eeq
with $dh_t$ the so-called one-particle hamiltonian, $dh_t=\sqrt{J}\, \sum_{j=1}^{N} ( \ket{j+1}\bra{j}\, dW_t^j + \mathrm{h.c.}\big)$, or equivalently, with $G_{i;j}=\bra{i}G_t\ket{j}$ the matrix elements of the matrix $G_t$ (we suppress the index $t$ to lighten the notation),
\begin{align} \label{eq:dG}
dG_{i;j} & =-2J\,G_{i;j}dt+J(G_{i+1;i+1}+G_{i-1;i-1})\delta_{i;j}dt  \\
&~~~ +i\sqrt{J}\big(G_{i;j-1}d\overline{W}_{t}^{j-1}+G_{i;j+1}dW_{t}^{j}-G_{i-1;j}dW_{t}^{i-1}-G_{i+1;j}d\overline{W}_{t}^{i}\big)~,\nonumber 
\end{align}
This is a matrix valued stochastic differential equation (SDE). The matrix $G$ satisfies a closed set of equations. In particular, the mean density $\mathfrak{n}_j:=\mathbb{E}[G_{j;j}]$ evolves diffusively, $\partial_t\mathfrak{n}_j = J\, \Delta\mathfrak{n}_j$ with $\Delta$ the discrete Laplacian.

At large time, the matrix $G$ reaches a steady distribution, which we shall denote $\mathbb{E}_\infty[\cdot]$. Since it is unitary, the dynamics  \eqref{eq:G-time} is isospectral, and all quantities $N_k:=\Tr(G^k)$ are constants of motion fixed by the initial condition $G_0$. We set $N_k=\mathfrak{m}_kN$.
The steady measure depends on this data. To compute the moments of $G$ in the steady measure is a linear problem specified by eq.\eqref{eq:dG}.  Let us compute the first few.

For the mean, we find
\beq \label{eq:EG1}
\mathbb{E}_\infty[G_{i;j}] = \mathfrak{m_1} \, \delta_{i;j} ~. %
\eeq
This reflects :\\
(a) decoherence as the off-diagonal elements vanish in average at large time,\\
(b) equilibrium as the mean density profile is uniform along the chain. (Recall that $G_{i;i}$ is the quantum expectation value of the particle number at site $i$).

For the quadratic cumulants, we find (with $i\not=j$) \cite{Bauer2019Equilibrium},
\begin{subequations} \label{eq:EG2}
\begin{align}
& \mathbb{E}_\infty[|G_{i;i}|^2]^c =\frac{(\Delta_2\mathfrak{m})^2}{N+1} ~,\quad 
\mathbb{E}_\infty[G_{i;i}G_{j;j}]^c=- \frac{(\Delta_2\mathfrak{m})^2}{N^{2}-1}~,\\
& \mathbb{E}_\infty[|G_{i;j}|^2]^c =\frac{N(\Delta_2\mathfrak{m})^2}{N^{2}-1}\simeq \frac{(\Delta_2\mathfrak{m})^2}{N}~,
\end{align}
\end{subequations}
with $(\Delta_2\mathfrak{m})^2 :=\mathfrak{m}_2- \mathfrak{m}_1^2$. We thus learn that fluctuations of the density at coincident points are of order $1/\sqrt{N}$, as expected for diffusive systems, but at order $1/N$ at non coincident points. More importantly, we observe that, although vanishing in average, coherences, represented by the off-diagonal elements of $G$, are of order $1/\sqrt{N}$ (provided the $\mathfrak{m}_k$ scale as $O(N^0)$). That is~: decoherence is at work in average but there are non-zero fluctuating coherences, sub-leading with the system size.

Suppose that the system is initially prepared in a factorized state, so that its initial two point functions are diagonal, $\Tr(\rho_0c^\dag_i c_j) = \mathfrak{n}_i\, \delta_{i;j}$, with $ \mathfrak{n}_i$ the initial local particle density at site $i$. Then, $\mathfrak{m}_1=\sum_i \mathfrak{n}_i/N$ is the mean density, and $\mathfrak{m}_2=\sum_i \mathfrak{n}_i^2/N$, so that  $(\Delta_2\mathfrak{m})^2$ is the variance of the initial density\footnote{If the initial state is not factorized, there are possible corrections to $\mathfrak{m}_2$ due to initial coherences.}. Eqs.\eqref{eq:EG2} tell us that, despite the absence of initial coherences, these are progressively, randomly, generated by the hopping process \eqref{eq:def-dH} and they survive the long time limit if the initial density is inhomogeneous. Phrased differently~: fluctuating coherences are dynamically produced from the inhomogeneities of the density profile. 

Higher cumulants can be recursively computed. Assuming that the initial state is such that all $\mathfrak{m}_k=\Tr(G_0^k)/N$ scale as $O(N^0)$ for large $N$, all cumulants of order $P$ scale as $1/N^{P-1}$. Let $W[A]:=\log Z[A]$ with $Z[A]:=\mathbb{E}_\infty\big[e^{\Tr(AG)}\big]$, with $A$ a generic test matrix, be their generating function. We have, order by order in power of $A$ (recall that $(\Delta_2\mathfrak{m})^2=\mathfrak{m}_2-\mathfrak{m}_1^2$) ~:
\beq \label{eq:W-extensive}
W(A) =  \mathfrak{m}_1\, \Tr A + \frac{1}{2}\frac{N(\Delta_2\mathfrak{m})^2}{N^2-1}\,\Tr A^2 - \frac{1}{2}\frac{(\Delta_2\mathfrak{m})^2}{N^2-1}\, (\Tr A)^2 + O(|\!|A|\!|^3) ~.
\eeq
This formula has a simple interpretation~: To leading order in the system size, the matrix of  two point functions converges to the non random, uniform, matrix $G_\mathrm{eq}=\mathfrak{m}_1 \mathbb{I}$, proportional to the identity, reflecting convergence toward equilibrium. There are sub-leading fluctuations, so that we can write,
\beq
 G \simeq G_\mathrm{eq} + \delta G~,\quad G_\mathrm{eq}=\mathfrak{m}_1 \mathbb{I} ~,\quad  \delta G\simeq O(1/\sqrt{N}) ~,
\eeq
where the fluctuations $\delta G$ decrease with the system size. Alternatively, the system density matrix converges to the equilibrium Gibbs state, up to fluctuations,
\beq
\rho \simeq \rho_\mathrm{eq} + \delta \rho~,\quad \rho_\mathrm{eq}=\frac{1}{Z_\mathrm{eq}}e^{-\mu\, \hat N_\mathrm{tot}}~,\quad  \delta \rho\simeq O(1/\sqrt{N}) ~,
\eeq
with $\hat N_{\mathrm{tot}}=\sum_j c^\dag_jc_j$ the total particle number, $\mu$ the chemical potential, related to the density via $\mathfrak{m}_1=1/(1+e^{\mu})$ and $Z_\mathrm{eq}$ the equilibrium partition function, $Z_\mathrm{eq}=(1+e^{-\mu})^N$. Recall that the number of particles is the only conserved quantity under the dynamics \eqref{eq:def-dH} (for any fixed realization of the noise). 

Hence, for large system sizes, the steady measure $\mathbb{E}_\infty$ is peaked around the equilibrium density matrix $\rho_\mathrm{eq}$, with fluctuations $\delta\rho$ sub-leading in the system size. It is interesting to compare with eq.\eqref{eq:rho-infty}. Q-SSEP gives access to these rare sub-leading fluctuations and yields to a precise description of their statistics.

\subsection{Closed Q-SSEP~: Where are we~?}
\label{sec:closed-details}

\hskip 0.6 truecm
$\bullet$ {\it Steady fluctuations and large deviation}

The closed Q-SSEP is actually ergodic, in any sector with a fixed number of particles, in the sense that the steady measure $\mathbb{E}_\infty$ is $SU(N)$ invariant. This follows from the fact that iterated products of the form
\[ e^{-idh_{t_1}}\, e^{-idh_{t_2}}\cdots e^{-idh_{t_n}} ~, \]
with $dh_{t_k}$ the one-particle hamiltonian increment \eqref{eq:G-time}, for any collection of time increments $dt_k$, visit densely the group $SU(N)$ (because the matrices $E_{j;j+1}:=\ket{j}\bra{j+1}$ and their adjoints form a system of simple root generators for $su(N)$). The dynamics generated by successive iterations of the infinitesimal group elements $e^{-idh_t}$ is thus ergodic enough to cover the group $SU(N)$, and this implies that the generating function, $Z[A]:=\mathbb{E}_\infty\big[e^{\Tr(AG)}\big]$, is $SU(N)$ invariant~: $Z[UAU^\dag]=Z[A]$ for any $U\in SU(N)$. As a consequence, it admits an integral representation, 
\beq \label{eq:Z-HIZ}
 Z[A]= \int\!\! dV\, e^{ \Tr(  AV^\dag G_0 V) } ~,
\eeq
where the integral is taken over the unitary group $U(N)$ with the invariant Haar measure.
This integral is known as  the Harish-Chandra-Itzykson-Zuber (HC-IZ) integral in random matrix theory (RMT) \cite{HarishChandra1957DifferentialOO,Itzykson1980ThePA}.

The proof is simple, using the $SU(N)$ invariance and the iso-spectral property of the Q-SSEP flow~:
\beqs \label{eq:IZ-trick}
Z[A] =  \int \!\! dV\,Z[VAV^\dag] = \mathbb{E}_\infty\Big[ \! \int \!\! dV\, e^{\Tr( AV^\dag GV)}\Big] = \int \!\! dV\,e^{\Tr( AV^\dag G_0 V)} ~.
\eeqs
In the first equality we use the $SU(N)$ invariance, in the second we insert the definition of $Z$ as the generating function for $\mathbb{E}_\infty$ and we permute integration and averaging, and in the last we use the property that the spectrum of $G$ is conserved by the flow, and thus non random, so that $\int \! dV e^{\Tr(AV^\dag G_0 V)}$ can be pulled out from the expectation with respect to $\mathbb{E}_\infty$. 

Alternatively, the system Hilbert, which is the Fock space of $N$ fermions, decomposes into the sum of sectors $\Lambda_M$ of fixed particle number, say $M=0,\cdots,N$~:
\[ 
\mathcal{H}_\mathrm{sys} = \Lambda_0\oplus\Lambda_1\oplus\cdots\oplus\Lambda_N ~,
\]
Each of the subspaces $\Lambda_M$ form a $U(N)$ irreducible representation (isomorphic to rank $M$ antisymmetric tensors in dimension $N$). Since, for each realisation of the noise, the Q-SSEP dynamics preserves the total number of particles, each of those sectors is stable under the Q-SSEP dynamics, and the invariant measure on each of them is that induced by the $SU(N)$ Haar measure. This is the measure we would get if we were sampling the states uniformly, according to the Haar measure, in the corresponding $SU(N)$ orbit. Thus, the system is ergodic in each sector with a fixed number of particles.

The generating function $Z[A]$, represented as a HC-IZ integral, can be analysed using tools from RMT. It depends on the spectrum of $G_0$ and hence on $\mathfrak{m}_k:=\Tr G_0^k/N$. Assuming that $\mathfrak{m}_k=O(N^0)$ for all $k$, then $w[A]:=\lim_{N\to\infty} \frac{1}{N} \log Z[NA]$ is finite, order by order in power of $A$, or alternatively, 
\beq \label{eq:W-dev}
\mathbb{E}_\infty\big[ e^{N\, \Tr(AG)} \big] \asymp_{N\to\infty} e^{N\,w[A]} ~.
\eeq 
From eq.\eqref{eq:W-extensive}, the first few terms are
\beq \label{eq:wz}
w[A] = \mathfrak{m}_1\, \Tr A + \frac{1}{2}\,(\Delta_2\mathfrak{m})^2\,\Tr A^2 + \frac{1}{3}(\Delta_3\mathfrak{m})^3\, \Tr A^3 + O(|\!|A|\!|^4) ~,
\eeq
with $(\Delta_2\mathfrak{m})^2=\mathfrak{m}_2-\mathfrak{m}_1^2$ and $(\Delta_3\mathfrak{m})^3:=\mathfrak{m}_3-3 \mathfrak{m}_2 \mathfrak{m}_1+2 \mathfrak{m}_1^3$.
This expansion suggests that $w[A]$ can be written as a series expansion in $\Tr A^k$ only~: $w[A]=\sum_k \frac{1}{k} \mathfrak{f}_k \Tr A^k$, where $\mathfrak{f}_k$ stands for the large $N$ limit of $N^{(k-1)} \mathbb{E}_\infty[G_{i_1i_2}G_{i_2i_3}\cdots G_{i_ki_1}]$ with $i_1,\cdots,i_k$ all distincts. This type of expectations, for connected products of $G$'s with indices $i_k$ placed along a loop, are going be also important in the study of the open Q-SSEP, since they are going to be the dominating ones. Intuitively, this indicates the emergence of a thermodynamic limit in which correlation functions factorize on dominating connected components.

Eq.\eqref{eq:W-dev} equivalently means that the steady probability distribution for $G$ satisfies a large deviation principle (compare with eq.\eqref{eq:large-dev}\footnote{To compare with eq.\eqref{eq:large-dev}, note that if $a_0$ is the mesh of the lattice and $L$ the size of the system, then $L=Na_0$ and $\varepsilon:=a_0/L=1/N$.} with $\varepsilon=1/N$), 
\beq \label{eq:largedev}
\mathbb{P}\mathrm{rob}_{\infty}[G=g]\asymp_{N\to\infty}e^{-N\,I(g)} ~,
\eeq
with rate function $I(g)$, the Legendre transform of $w[A]$. Its series expansion around its minimum reads :
\beq
I(g)=\frac{\Tr(g-\mathfrak{m}_{1}\mathbb{I})^{2}}{2(\Delta_2\mathfrak{m})^2}-\frac{\Tr(g-\mathfrak{m}_{1}\mathbb{I})^{3}}{3(\Delta_2\mathfrak{m})^{3}}\frac{(\Delta_3\mathfrak{m})^3}{(\Delta_2\mathfrak{m})^{3}} +O(|\!|g-\mathfrak{m}_{1}\mathbb{I}|\!|^4) ~.
\eeq
To leading order, $G$ is Gaussian with mean $\mathfrak{m}_{1}\mathbb{I}$ and variance of order $(\Delta_2\mathfrak{m})/N^{1/2}$. But they are higher cumulants. 
Clearly, there is a need for a better, more complete, non perturbative, description of this large deviation function (however see \cite{Guionnet2002large}).

\medskip

$\bullet$ {\it Fluctuation dynamics}

The time evolution of fluctuations is governed by the stochastic equation \eqref{eq:motion-rho} for the density matrix, or eq.\eqref{eq:dG} for the two-point functions. For the average density matrix, $\bar \rho_t=\mathbb{E}[\rho_t]$, it leads to the Lindblad equation eq.\eqref{eq:L1-dyn}. For higher moments, say $\mathbb{E}[\rho_t\otimes\cdots\otimes\rho_t]$, made of products of $R$ replicas of the density matrix, the dynamics is also given by a Lindblad equation,
\beq \label{eq:LR-dyn}
\partial_t \mathbb{E}[\rho_t\otimes\cdots\otimes\rho_t] = \mathcal{L}^{(R)}_{\mathrm{ssep}}(\mathbb{E}[\rho_t\otimes\cdots\otimes\rho_t]) ~,
\eeq
with a Lindbladian $\mathcal{L}^{(R)}_{\mathrm{ssep}}$ similar to that for one replica, as in eq.\eqref{eq:L-edge}, but with the jump operators $\ell^\pm_j$ replaced by their sum over the various replicas, $\ell^\pm_{j} \to\sum_a \ell^\pm_{a;j}$, where the index $a$ labels the replicas and the individual operators $\ell^\pm_{a;j}$ only acts on the $a$-th replica.

As shown in \cite{Bernard2021Dynamics}, eq.\eqref{eq:LR-dyn} can be mapped to a spin chain dynamics, with underlying $gl(2R)$ symmetry algebra. Namely, the Lindbladian $\mathcal{L}^{(R)}_{\mathrm{ssep}}$ can be written as follows,
\beq \label{eq:L-tensorC-GL}
\mathcal{L}^{(R)}_{\mathrm{ssep}}  = J \sum_{j=1}^N \Big( \sum_{A,B} G_{j+1}^{AB}G_j^{BA} - \frac{1}{2}(C_{j+1}+C_j+2R) \Big) ~,
\eeq 
where the $G_j^{AB}$'s, with $A=1,\cdots,2R$, form a set of super-operators (i.e. linear maps acting on operators), acting locally on site $j$, with commutation relations,
\beq \label{eq:G-relation}
 \big[G_j^{AB},G_k^{CD}\big]=\delta_{j,k}(\delta^{BC}G_j^{AD}-\delta^{DA}G_j^{CB}) ~,
\eeq
and $C_j+R:=\sum_AG_j^{AA}$. The super-operators $G_j^{AB}$ act on any operator $X$ by left or right multiplications of $X$ by fermionic creation or annihilation operators. They form a representation of the Lie algebra $gl(2R)$ on the space of operators localized at site $j$, which is is isomorphic to the exterior algebra in dimension $2R$. This operator space is also isomorphic to the direct sum of all fundamental representations of $sl(2R)$. Thus, the Lindblad dynamics \eqref{eq:L-tensorC-GL} is that of  a spin chain with local degree of freedom in the fundamental representations of $sl(2R)$.

This observation allows to borrow information and techniques from spin chain studies. In particular, it allows to study steady or low lying states. The integrability for one replica has been shown in \cite{Essler2020Integrability} and in some higher replica sectors. The question whether it is integrable in all sectors, for any number of replicas, is still open \cite{Bernard2021Dynamics}. Lindblad integrable dynamics have recently been analysed \cite{eisler2011crossover,Medvedyeva2016Exact,Naoyuki2019dissipative,Ribeiro2019integrable,nakagawa2020exact,buca2020dissipative,ziolkowska2020yang,deLeeuw2021Constructing,robertson2021exact} but finding a series of integrable Lindbladians coding for all replicas dynamics is still an open question.

It also allows to study the dynamics in the continuous scaling limit. Let $a_0$ be the mesh of lattice, i.e. the size of the elementary edges, so that the total length of the chain is $L=Na_0$. The scaling limit is the limit,
\beq \label{eq:scaling-lim}
a_0\to 0,\ N\to\infty,\ J\to\infty,\quad \mathrm{with}\ L:=a_0N,\ D:=Ja_0^2,\ x:=ja_0~\ \mathrm{fixed},
\eeq
with $J$ the coupling constant in eq.\eqref{eq:def-dH} and $j$ the position of the insertion point. This amounts to impose a diffusive scaling relation between time and space. $D$ has the dimension of diffusion constant. We set $\varepsilon:=a_0/L=1/N$, as in MFT.

The discrete diffusion equation for the local particle number has a simple scaling. Let $\mathfrak{n}(x,t)$ be the scaling limit of the density, $\mathfrak{n}(x,t)=\mathfrak{n}_j(t)$ for $\mathfrak{n}_j(t)=\mathbb{E}[G_{jj}(t)]$.  As a consequence of eq.\eqref{eq:dG}, it evolves diffusively,
\beq \label{eq:diff-n}
 \partial_t \mathfrak{n}(x,t) = {D}\,\nabla_x^2\, \mathfrak{n}(x,t) ~,
\eeq
as expected since, in average, the Q-SSEP dynamics is purely diffusive (by construction). At large time, $\mathfrak{n}(x,t)$ reach its uniform steady value $\mathfrak{n}=\mathfrak{m}_1$.

Let us now look at the dynamics of the quadratic fluctuations in the scaling limit. We follow the analysis performed in \cite{Bernard2021Dynamics} and define the scaling limits of the density correlations and coherence fluctuations~: 
\begin{subequations}
\begin{align}
\mathfrak{g}_+(x,y,t) &:= \lim_{\mathrm{scaling}}\, \mathbb{E}[ G_{jj}(t)G_{kk}(t)]~,\\
\mathfrak{g}_-(x,y,t) &:= \lim_{\mathrm{scaling}}\, \mathbb{E}[ G_{jk}(t)G_{kj}(t)] ~.
\end{align}
\end{subequations}
They admit an $\varepsilon$-expansion:
\beq \label{eq:ep-g-expand}
\mathfrak{g}_\sigma(x,y,t) = \mathfrak{g}^{(0)}_\sigma(x,y,t) + \varepsilon\, \mathfrak{g}^{(1)}_\sigma(x,y,t) + O(\varepsilon^2) ~.
\eeq
They satisfy a hierarchy of linear evolution equations, with a triangular structure.
The leading terms  $\mathfrak{g}^{(0)}_\sigma(x,y,t)$ evolve diffusively,
\beq \label{eq:g-diff}
\partial_t \mathfrak{g}^{(0)}_\sigma(x,y,t) = {D}(\nabla_x^2+\nabla_y^2) \mathfrak{g}^{(0)}_\sigma(x,y,t) ~.
\eeq
In absence of long range order in the initial state, the off-diagonal coherences initially vanish at leading order in the scaling limit, while the density correlations are initially factorized. Since they are diffusively transported, this holds true at any time, and
\beq \label{eq:g+-leading}
\mathfrak{g}^{(0)}_+(x,y,t)= \mathfrak{n}(x,t)\mathfrak{n}(y,t)~,\quad \mathfrak{g}^{(0)}_-(x,y,t)=0 ~.
\eeq
The first relation expresses the fact that density fluctuations are sub-leading in the system size, so that the density is self-averaging at leading order as expected (cf. the above discussion of the steady measure). The second relation echoes that coherences are also sub-leading in the system size as imposed by decoherence. These properties can be checked on their large time steady values, since eqs.(\ref{eq:EG1},\ref{eq:EG2}) read $\mathbb{E}_\infty[G_{jj}G_{kk}]= \mathfrak{m}_1^2+\frac{1}{N}(\Delta_2 \mathfrak{m})^2\delta_{jk}$ and $\mathbb{E}_\infty[G_{jk}G_{kj}]= \mathfrak{m}_1^2\delta_{jk} +\frac{1}{N}(\Delta_2 \mathfrak{m})^2$ with $\mathfrak{m}_1=\mathfrak{n}$. Furthermore, if the average densities are initially uncorrelated\footnote{which is always the case if the system is initially prepared in a given state}, then $\mathfrak{g}^{(1)}_+(x,y,t)=0$, so that the density cumulants are actually of second order in $\varepsilon$.  

Remarkably, as a consequence of eq.\eqref{eq:dG}, the time evolution of the (sub-leading) coherence fluctuations is essentially diffusive but sourced by the density in-homogeneities \cite{Bernard2021Dynamics},
\beqa \label{eq:g1-evolution}
&& \partial_t \mathfrak{g}^{(1)}_-(x,y,t) = {D}(\nabla_x^2+\nabla_y^2) \mathfrak{g}^{(1)}_-(x,y,t) + \mathfrak{h}_+(x,y,t) ~,\\
&& \mathrm{with}\ h_+(x,y,t)= 2\,{D}{L}\, \nabla_x\nabla_y\big(\delta(x-y)\mathfrak{n}(x,t)\mathfrak{n}(y,t)\big) ~. \nonumber
\eeqa
Higher order correlations satisfy similar equations.

\begin{figure}
\begin{minipage}{.45\textwidth}
\includegraphics[width=0.9\textwidth]{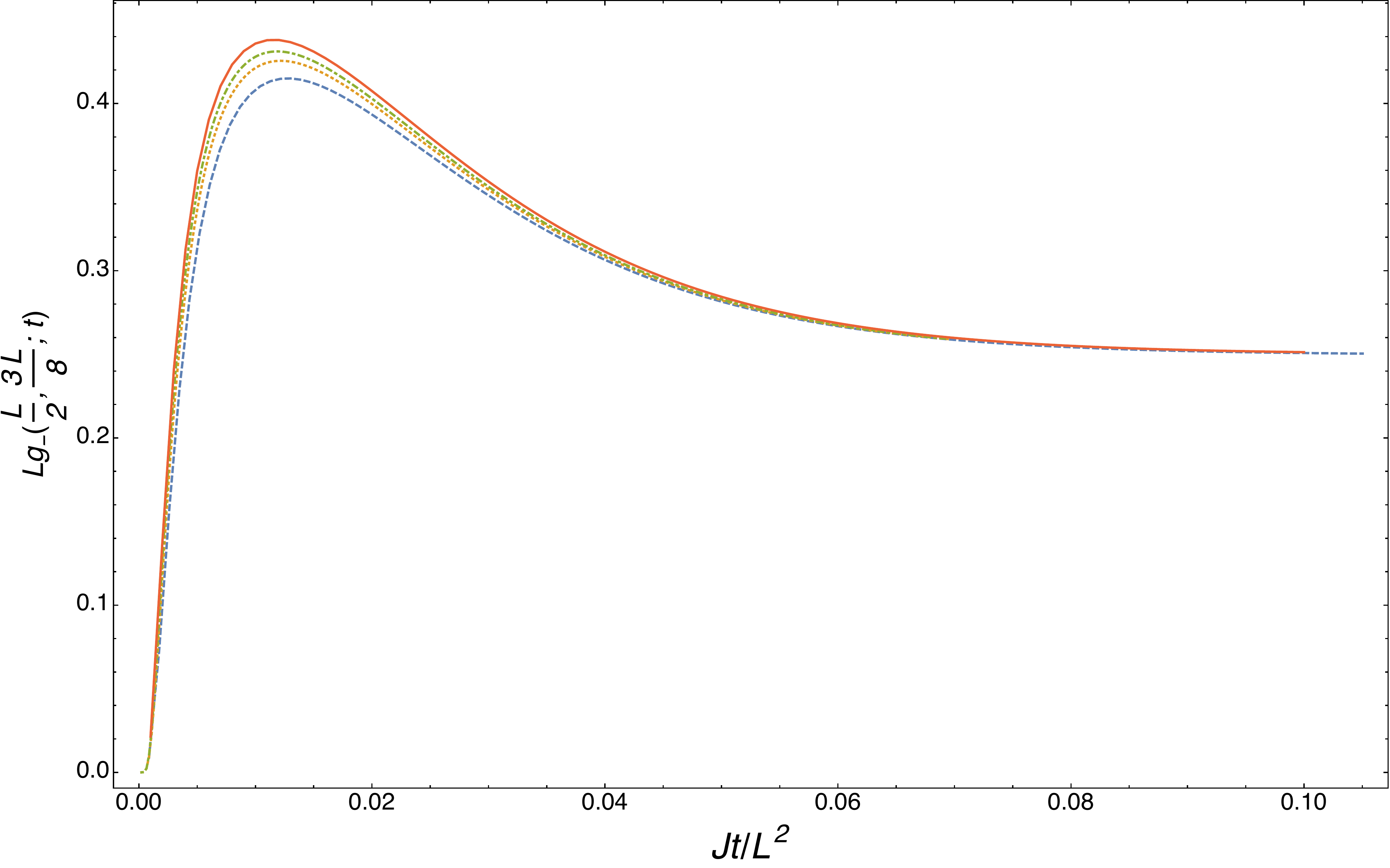}
\end{minipage}\hfill
\begin{minipage}{.45\textwidth}
\includegraphics[width=0.9\textwidth]{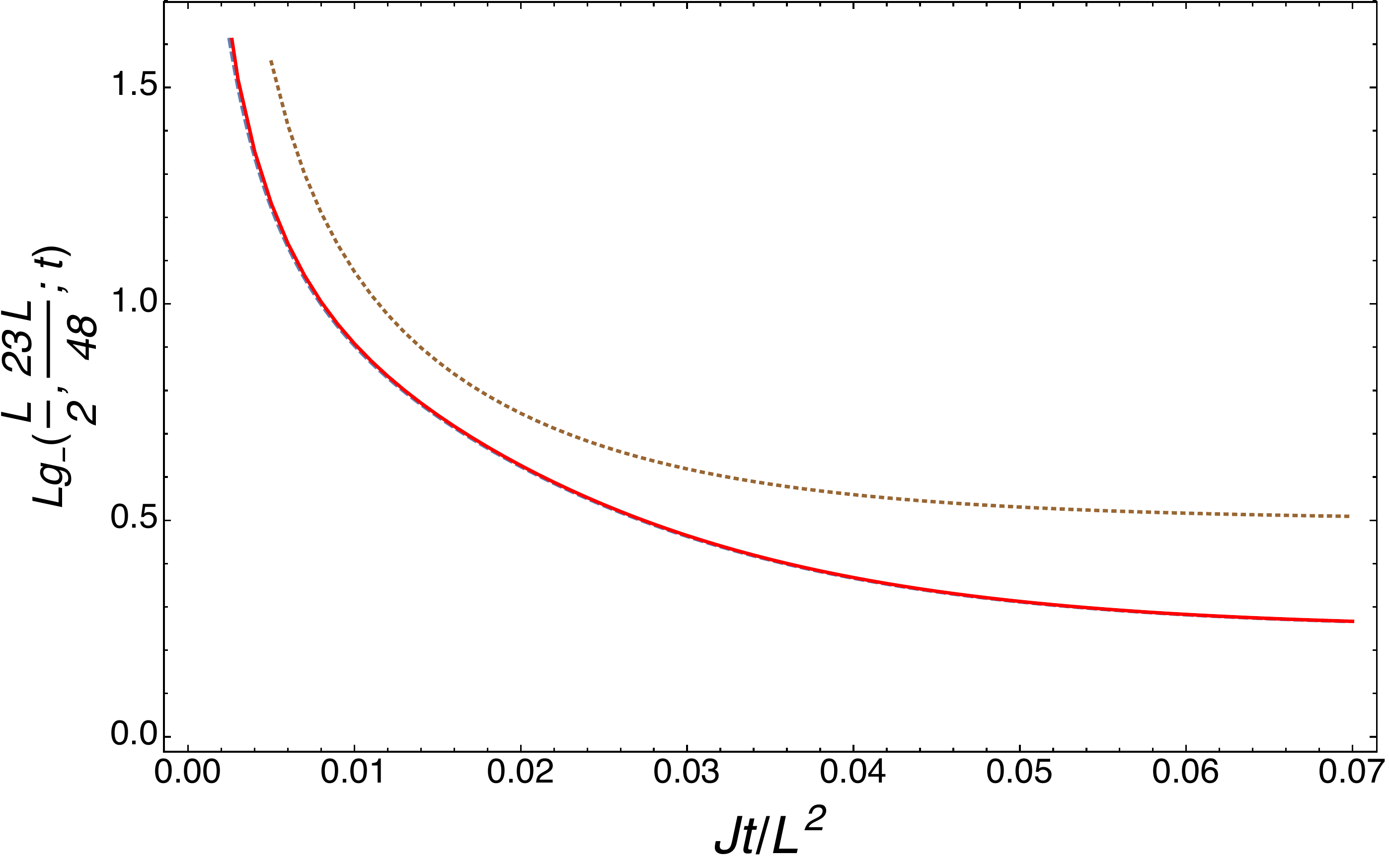}
\end{minipage}
\caption{\it [Left]: Comparison between $g_{-}(L/2,3L/8;t)$ as a function of
$Jt/L^2$ on a ring of $L=48$ (blue dashed), $L=96$ (yellow dotted),
  $L=192$ (gree dot-dashed) sites with domain wall initial conditions
  and $\mathfrak{g}^{(1)}_-(x,y,t)$ (red solid). The continuum limit result is approached throughout the depicted time range as $L$ increases. [Right]: Comparison between $g_{-}(L/2,23L/48;t)$ as a function of $Jt/L^2$ on a ring of $L=96$ (blue dashed) sites with domain wall initial conditions and $\mathfrak{g}^{(1)}_-(x,y,t)$ (red solid). The dotted drown line is the result of pure diffusion only, without the source term $\mathfrak{h}_+(x,y,t)$. Figures taken from \cite{Bernard2021Dynamics}.}
\label{fig:latt-vs-cont}
\end{figure}

The validity of the scaling evolution equations (\ref{eq:diff-n},\ref{eq:g-diff},\ref{eq:g+-leading},\ref{eq:g1-evolution}) over a significant range of time and space scales has been numerically checked in \cite{Bernard2021Dynamics}, for various initial configurations. In particular, these numerical results show that the source term $\mathfrak{h}_+$ in eq.\eqref{eq:g1-evolution} is relevant in this scaling range (so that the evolution of the coherences is not purely diffusive but modified by the interaction reflected in local density in-homogeneities). See Figure \ref{fig:latt-vs-cont}.

The structure of the evolution equations (\ref{eq:diff-n},\ref{eq:g-diff},\ref{eq:g+-leading},\ref{eq:g1-evolution}) is similar to that of the fundamental MFT equations \eqref{eq:mft}. In both cases, the evolution is mainly diffusive and non-random at leading order, as dictated by the Fourier-Fick's law, but with relevant corrections for fluctuations at sub-leading order in $\varepsilon$. In MFT these corrections are encoded into the Langevin equation \eqref{eq:mft}. It would be interesting to know,  in the case of Q-SSEP, whether there exits an effective stochastic process or an effective field theory describing them. If the answer is positive, this would provide the first step in constructing the Quantum Mesoscopic Fluctuation tTheory (Q-MFT).

\medskip

$\bullet$ {\it Entanglement statistics and dynamics}

Exact results on the statistics of entanglement can be obtained by borrowing techniques from RMT. Let us assume that the system is initially prepared in a state $\ket{\psi_M}$ with $M$ particles, with density $\mathfrak{n}=M/N$. Since purity and  particle number are conserved by the Q-SSEP dynamics, it will remain in a pure state $\ket{\psi_M(t)}$ with $M$ particles at any time. Following \cite{Bernard2021entanglement}, let us consider the entanglement of a subsystem $\mathcal{A}_{\ell}=\{1,\ldots ,\ell\}$, as measured by the R\'enyi-$q$ entropies, 
\beq
S_{q}(t) :=(1-q)^{-1}\log\Tr \rho_\ell(t)^q ~,
\eeq
 where $\rho_\ell$ is the system density matrix reduced to $\mathcal{A}_{\ell}$~: $\rho_\ell(t) :=\Tr_{N\setminus\ell}(|\psi_M(t)\rangle\langle \psi_M(t)|)$. We assume $\mathcal{A}_{\ell}$ to be extensive and set  $\xi=\ell/N$. The R\'enyi entropies are then typically extensive $S_q\sim \ell\, s$, with $0\leq s\leq \log 2$. We are interested in their statistics at large time. 
 
As shown in \cite{Bernard2021entanglement}, in the large size limit, ($N$,  $\ell$, $M$ large with fixed ratios  $\xi=\ell/N$, $\mathfrak{n}=M/N$), the R\'enyi entropies satisfy a large deviation principle, w.r.t. the invariant measure $\mathbb{E}_\infty$, 
\beq
\mathbb{P}\mathrm{rob}_\infty[S_q=\ell\, s] \asymp_{N\to\infty} e^{-\ell^2\, I_q(s)} ~,
\eeq
with a rate function $I_q(s)$, which depends on the initial density $\mathfrak{n}$ and on the fraction  of volume $\xi$ occupied by the sub-system. Compare with eq.\eqref{eq:large-dev}.

\begin{figure}
\begin{minipage}{.45\textwidth}
\includegraphics[width=0.9\textwidth]{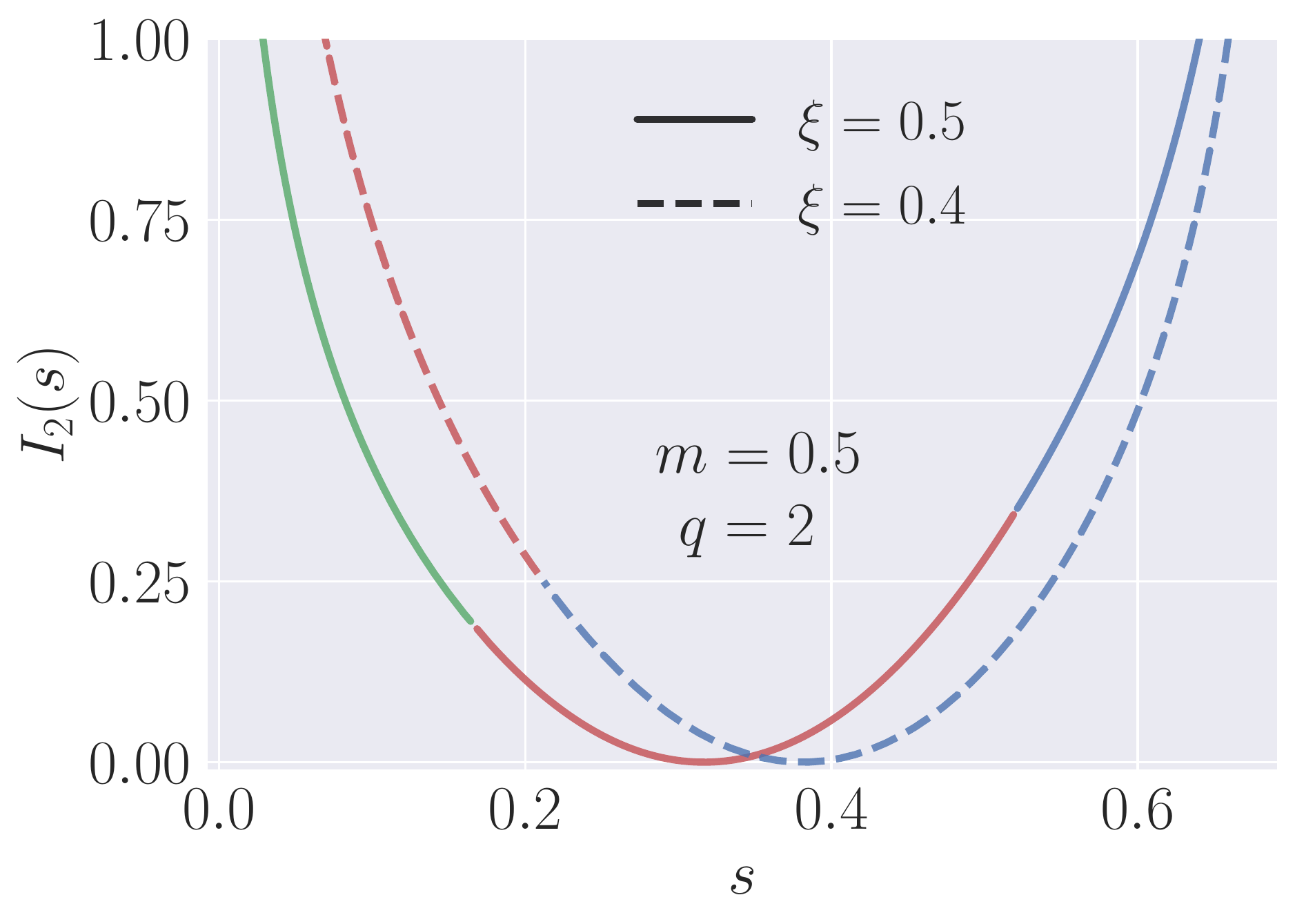}
\end{minipage}\hfill
\begin{minipage}{.45\textwidth}
\includegraphics[width=0.9\textwidth]{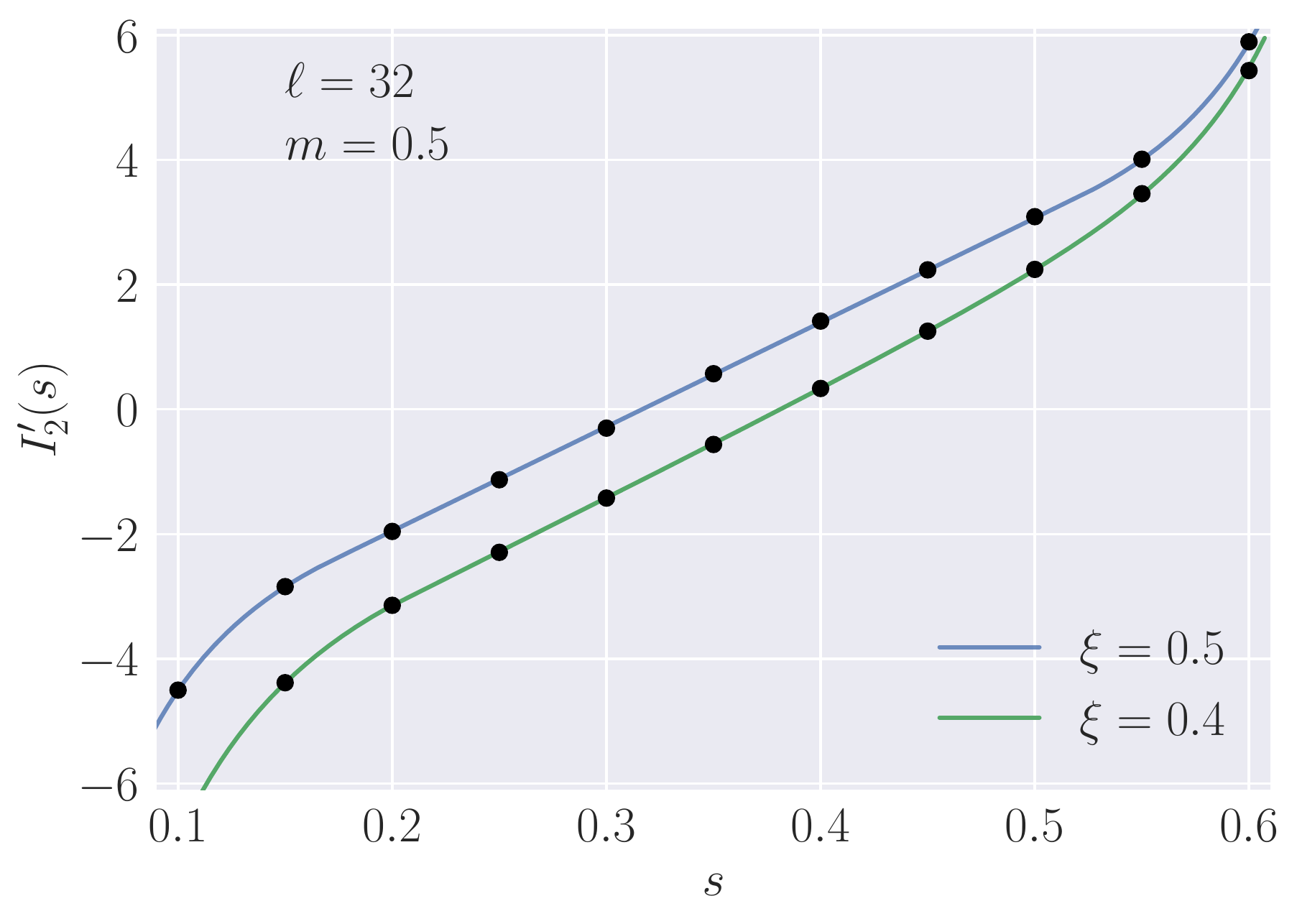}
\end{minipage}
\caption{\it [Left]: Rate function for the R\'enyi-$2$ entropy, for $m=1/2$, and different values of $\xi$. For $\xi<m$ and $\xi=m$ respectively, two and three phases appear, which correspond to different colors. [Right]: Analytic predictions for the derivative $I^\prime_2(s)$ (solid lines), against numerical data from Monte Carlo simulations for $\ell=32$, $L=\ell/\xi$ and $M=L/2$ (dots). The numerical error is not visible at the scales of the plot. Figures taken from \cite{Bernard2021entanglement}.}
\label{fig:qssep-renyi}
\end{figure}

This rate function can be evaluated using RMT techniques (since the invariant measure on the space of fixed number of particles is that induced by the Haar measure on $U(N)$). The computation is mapped to an extremization problem. It takes three different analytic forms in three different regions~\cite{Bernard2021entanglement}:
\begin{itemize}
\item For low values of the entropy, $0<s<s_-$ for some $s_-$ depending on $\xi$ and $\mathfrak{n}$, the rate function is dominated by contributions from pure states, and  diverges when $s\to 0^+$. In particular, $I_{q=2}(s)= -\frac{1}{2}\log s + O(s)$ as $s\to 0^+$.
\item The intermediate region, $s_-<s<s_+$, contains the minimum $\bar s$ of the rate function, which is the most-probable value of the entropy. In particular, for $q=2$ it is Gaussian, $I_{q=2}(s)=\frac{1}{2\gamma}(s-\bar s)^2$ with $\gamma$ a computable constant.
\item For high values of the entropy, $s_+<s<s_\mathrm{max}=\log 2$ for some $s_+$, the rate function is dominated by contributions from maximally mixed states, and  diverges when $s\to s_\mathrm{max}$. In particular, $I_{q=2}(s)= -\frac{1}{2}\log |s-s_\mathrm{max}| + O(s-s_\mathrm{max})$ as $s\to s_\mathrm{max}^-$.
\end{itemize}
The transitions between these different regimes are non-analytic, signalling phase transitions. See Figure \ref{fig:qssep-renyi}.

Clearly, it is worth understanding the entanglement dynamics and its scaling limit. The nature of the spreading of entanglement in this class of models is yet unknown. On one hand, local quantities spread diffusively, with possible sub-leading relevant corrections, see eqs.(\ref{eq:diff-n},\ref{eq:g+-leading},\ref{eq:g1-evolution}). One the other hand, the numerical analysis of \cite{Knap2018Entanglement} done on noisy spin chains (which are noisy many-body models, similar to Q-SSEP while not identical) suggests dynamical scaling exponents compatible with those of the Kadar-Parisi-Zhang (KPZ) class.
Comparing the resulting hydrodynamics description of entanglement spreading with the membrane picture \cite{Mezei2018membrane,Zhou2019emergent,Gullans2019entanglement,Agon2019bit,Zhou2020entanglement} which recently emerged from studies of random unitary circuits will also be interesting. This will provide another step toward the construction of a Q-MFT. We hope to report soon on this problem \cite{BPMJInPreparationQSSEP}.

\subsection{Open Q-SSEP~: Fluctuations out-of-equilibrium}

To simplify the notation, we now set the system length to unity, $L=1$. Thus, the lattice mesh is $a_0=1/N$ and the small dimensionless scaling parameter is $\varepsilon :=a_0/L=1/N$.

Contrary to the closed version, the open Q-SSEP involves injection and extraction processes at the two ends of the system interval, as specified in eq.\eqref{eq:Q-flow}. Due to the boundary processes, this evolution is not unitary. Nevertheless, for each realisation of the noise, the model is still quadratic and can be analysed by studying its two-point functions, $G_{i;j}=\Tr(\rho_t c_j^\dag c_i)$. The latter satisfy the following stochastic differential equation (SDE)~:
\begin{align} \label{eq:dG-open}
dG_{i;j} & =-2J\,G_{i;j}dt+J(G_{i+1;i+1}+G_{i-1;i-1})\delta_{i;j}dt  \\
&~~~~ +i\sqrt{J}\big(G_{i;j-1}d\overline{W}_{t}^{j-1}+G_{i;j+1}dW_{t}^{j}-G_{i-1;j}dW_{t}^{i-1}-G_{i+1;j}d\overline{W}_{t}^{i}\big) \nonumber \\
&~~~~~ + \sum_{p\in\{1,N\}}\big(\alpha_p\delta_{i,p}\delta_{i;j} - \frac{1}{2}(\alpha_{p}+\beta_{p})(\delta_{i,p}+\delta_{j,p})G_{i;j}\big)dt ~, \nonumber
\end{align}
Compared to eq.\eqref{eq:dG} valid in the closed set-up, eq.\eqref{eq:dG-open} contains boundary terms reflecting the injection and extraction processes with rates $\alpha_{1,N}$ and $\beta_{1,N}$, as illustrated in the evolution equation for the mean occupation numbers, ${n}_j:=\mathbb{E}[G_{j;j}]$~:
\beq \label{eq:n-open}
\partial_t{n}_j= J \Delta {n}_j + \hskip -0.3truecm \sum_{p\in\{1,N\}}\hskip -0.2truecm \delta_{j;p}\big(\alpha_{p}(1-{n}_p)-\beta_{p}{n}_p\big) ~,
\eeq
with $\Delta$ the discrete Laplacian.

In average, and in the diagonal sector (diagonal w.r.t. the particle number operators), the open Q-SSEP dynamics reproduce the open classical SSEP with injection and extraction of particles at its two ends.

As for the closed case, the two-point functions reach a steady distribution at large time. Since eq.\eqref{eq:dG-open} is linear, moments of $G$ can be recursively computed, at least for the first of them. For the mean, from eq.\eqref{eq:n-open} we get,
\beq
\mathbb{E}_\infty[G_{i;j}] = \frac{n_a(N+b-j) + n_b(j+a-1)}{N+b+a-1}\, \delta_{i;j} ~,
\eeq
with $n_{a}:=\frac{\alpha_{1}}{\alpha_{1}+\beta_{1}}$, $n_{b}:=\frac{\alpha_{N}}{\alpha_{N}+\beta_{N}}$, $a:=\frac{J}{\alpha_{1}+\beta_{1}}$, $b:=\frac{J}{\alpha_{N}+\beta_{N}}$. In the scaling limit, $N\to\infty$ at $x=j/N$ fixed (with $n_a, n_b, a, b$ fixed), the average profile becomes,
\beq
n^*(x)= n_a + x(n_b-n_a) ~.
\eeq
It interpolates linearly between the two boundary occupations $n_a$ and $n_b$. The average current, which is proportional to the gradient of the density by the Fourier-Fock's, is thus non-vanishing, provided the two boundary occupations are un-balanced. The system is hence out-of-equilibrium.

Higher moments of the two-point functions, w.r.t. the steady measure $\mathbb{E}_\infty$, were analysed in \cite{Bernard2019Open,Bernard2021Solution}. It has been shown that, in the large size limit, the leading multi-point cumulants of order $P$, say $\mathbb{E}_\infty[G_{i_1j_1}\cdots G_{i_Pj_P}]^c$, come from the expectation values  of cyclic products $G_{i_1i_P}\cdots G_{i_3i_2}G_{i_2i_1}$. They scale proportionally to $1/N^{P-1}$ in the large $N$ limit. This may be compared to the analysis of the large deviation function in the closed case, see eqs.(\ref{eq:W-extensive},\ref{eq:wz}). Other cumulants of $G_{ji}$'s are sub-leading at large $N$. For instance, in the limit $N\to\infty$ with $x=i/N$, $y=j/N$ fixed, one has (for $x<y$), 
\begin{subequations} \label{eq:N=2-QC}
\begin{align} 
 \mathbb{E}_\infty[G_{ij} G_{ji}]^c&= \frac{1}{N}(\Delta n)^2\, x(1-y) +O(N^{-2}) ~,\\
 \mathbb{E}_\infty[G_{ii}^2]^c  &=  \frac{1}{N}(\Delta n)^2\, x(1-x) +O(N^{-2}) ~,\\
\mathbb{E}_\infty[G_{ii}G_{jj}]^c &=- \frac{1}{N^2}(\Delta n)^2\, x(1-y) +O(N^{-3}) ~,
\end{align}
\end{subequations}
with $\Delta n:= n_b-n_a$.
As in the closed case, quadratic coherence fluctuations scale as $O(1/{N})$ in the thermodynamic limit, while the quadratic density correlations at distinct positions scale as $O(1/N^2)$ and hence are sub-leading.

The statistics is however not Gaussian as it can be seen by computing the leading cumulants with three points (with $x<y<z$):
\beq
\mathbb{E}_\infty[G_{ik}G_{kj}G_{ij}]^c= \frac{1}{N^2} (\Delta n)^3\, x(1-2y)(1-z) +O(N^{-3}) ~,
\eeq
in the scaling limit $N\to\infty$ with $x=i/N$, $y=j/N$ and $z=k/N$ fixed. 

More generally, moments of cyclic products $G_{i_1i_P}\cdots G_{i_3i_2}G_{i_2i_1}$ depend on whether the ordering of the points $x_k=i_k/N$ along the chain and along the cycles defined by these  products coincide or not. Fixing an order along the chain interval, the different orderings along these cyclic products are indexed by single cycle permutations of $P$ elements. The rule for this correspondence is that by turning around the oriented loop indexed by a permutation $\sigma$ one successively encounters the points labeled as $i_{1}$, $i_{\sigma(1)}$, $i_{\sigma^{2}(1)}$, $\cdots$, up to closing the loop back to $i_{\sigma^{P}(1)}=i_1$. We can thus choose $0\leq x_1<\cdots<x_P \leq1$, with $x_k=i_k/N$, and index these cyclic cumulants by single cycle permutation $\sigma$. The claim is \cite{Bernard2019Open,Bernard2021Solution}~:
 \beq \label{eq:Q-loop}
\mathbb{E}[G_{i_1i_{\sigma^{P-1}(1)}}\cdots G_{i_{\sigma^2(1)}i_{\sigma(1)}}G_{i_{\sigma(1)}i_1}]^c =\frac{(\Delta n)^P}{N^{P-1}}\, g^{(P)}_\sigma(x_1,\cdots,x_P)+ O(\frac{1}{N^{P}}) ~,
\eeq
For instance, for $x_1<x_2<x_3<x_4$, we have :
\begin{align*}
\mathbb{E}_\infty[G_{i_1i_4}G_{i_4i_3}G_{i_3i_2}G_{i_2i_1}]^c &= N^{-3}(\Delta n)^4\, x_1(1-3x_2-2x_3+5x_2x_3)(1-x_4) +O(N^{-4}) ~,\\
\mathbb{E}_\infty[G_{i_1i_2}G_{i_2i_4}G_{i_4i_3}G_{i_3i_1}]^c &=N^{-3}(\Delta n)^4\, x_1(1-3x_2-2x_3+5x_2x_3)(1-x_4) +O(N^{-4}) ~,\\
\mathbb{E}_\infty[G_{i_1i_4}G_{i_4i_2}G_{i_2i_3}G_{i_3i_1}]^c &=N^{-3}(\Delta n)^4\, x_1(1-4x_2-x_3+5x_2x_3)(1-x_4) +O(N^{-4}) ~,
\end{align*}
with $x_k=i_k/N$. A few other cumulants can be directly computed. They are all polynomials of degree at most one in each of their variables.

Hence, although decoherence is at play in the average open Q-SSEP behaviour, fluctuating quantum coherences survive at large time, although sub-leading. Their statistics possesses a rich and intriguing structure. The scalings of their cumulants with the system size is such they satisfy a large deviation principle. The open Q-SSEP seems thus to be an appropriate quantum extension the open classical SSEP.

\subsection{Open Q-SSEP~: Where are we~?}

\hskip 0.6 truecm
$\bullet$ {\it Steady fluctuations and large deviation}

The scaling of these cumulants with $N$ signals the existence of a large deviation function, with small parameter $\varepsilon=1/N$ as for the MFT, see \eqref{eq:W-dev}. But, no explicit, or even implicit, formula for this large deviation function is presently known. In particular, it is not yet known whether this large deviation function derives from a minimization principle as its analogue \eqref{eq:Wh-int} in MFT does. The answer to this question would be an important step towards constructing a Q-MFT.

However, a procedure to compute all cyclic cumulants of the two-point functions, in the steady measure, has been proposed in \cite{Bernard2021Solution}. It is based on an interplay between permutations and polynomials and in introducing appropriate generating functions. Let us just present it in the simplest case of so-called regular cyclic cumulants for which the order of the points along the cycle and the chain coincide. Let $\omega_P(x_1,\cdot,x_P)$ be the corresponding scaled cumulants,
\beq
\mathbb{E}_\infty[G_{i_1i_P}\cdots G_{i_3i_2}G_{i_2i_1}]= N^{1-P}\, (\Delta n)^P\,\omega_P(x_1,\cdots,x_P) + O(N^{-P}) ~.
\eeq
with $x_k=i_k/N$ and $0 \leq x_1\leq \cdots \leq x_P \leq 1$. Let us then define generating functions $\mathcal{C}_k(z)$, made of multiple derivatives of these cumulants, via the formal power series,
\beq
\mathcal{C}_k(z) := \sum_{P\geq 0} z^{P}\ \nabla_{x_{P+1}}\cdots\nabla_{x_1}\omega_{P + k + 1}(x_1,\cdots, x_{P + k + 1}) ~,
\eeq
Since all $\omega_{P+k+1}$ are polynomials of degree at most one in each of its variables, the term proportional to $z^P$ in the function $\mathcal{C}_k(z)$ depends only on the remaining $k$ variables $x_{P+k+1},\cdots,x_{P+2}$ on which the derivatives are not acting. To stabilize the notation, we renamed them as $y_l:=x_{P+k+1-l}$ for $l=1,\cdots,k$, so that $\mathcal{C}_k(z)$ depend only on $y_0,\cdots,y_{k-1}$. Surprisingly, the conditions for the stationarity of the measure $\mathbb{E}_\infty$ can be recasted into the following recursion relation~:
\begin{equation} \label{eq:Ck-series} 
\mathcal{C}_{k+1}(z)= \Big[ \big(\mathfrak{c}(z) + \frac{y_k}{z}\big)\mathcal{C}_{k}(z)\Big]_+ ~,
\end{equation}
 where  $\big[\cdots\big]_+$ means the part of the Laurent series with positive degrees and $\mathfrak{c}(z):=\big(\sqrt{1+4z}-1\big)/2z=1-z+2z^2-5z^3+\cdots$ is the generating function of alternating Catalan numbers. This allows to compute them recursively, say  $\mathcal{C}_0(z)=\mathfrak{c}(z)$, $\mathcal{C}_1(z)=\mathfrak{c}(z)^2(1-y_0)$, $\mathcal{C}_2(z)=\big(\mathfrak{c}(z)^3+y_1z^{-1}(\mathfrak{c}(z)^2-1)\big)(1-y_0)$, etc.
 The cumulants $\omega_P$ are then recovered from the functions $\mathcal{C}_{P-1}(z)$ by evaluating them at $z=0$ and by renaming back the $y$'s  in terms of the $x$ positions~:
 \beq \label{eq:omega-Ck}
  \omega_{P}(x_1,x_2,\cdots) = x_1\,\mathcal{C}_{P-1}(0)\vert_{(y_0=x_{P},y_1=x_{P-1},\cdots,y_{P-1}=x_2)} ~.
\eeq
For instance,
\beqa \label{eq:C-exemple}
\mathcal{C}_3(0) &=& \! \big(5\,y_{1}\,y_{2} - 3\,y_{2} - 2\,y_{1} + 1\big)(1-y_0) ~,\nonumber\\
\mathcal{C}_4(0) &=& \! \big(- 14\,y_{1}\,y_{2}\,y_{3} + 9\,y_{2}\,y_{3} + 7\,y_{1}\,y_{3} + 5\,y_{1}\,y_{2}-4\,y_{3} - 3\,y_{2}-2\,y_{1} + 1\big)(1-y_0) ~, \\
\mathcal{C}_5(0) & =& \! \big(42\,y_{1}\,y_{2}\,y_{3}\,y_{4} - 28\,y_{2}\,y_{3}\,y_{4} - 23\,y_{1}\,y_{3}\,y_{4} - 19\,y_{1}\,y_{2}\,y_{4} - 14\,y_{1}\,y_{2}\,y_{3}  + 14\,y_{3}\,y_{4}   \nonumber\\ 
& & \! + 12\,y_{2}\,y_{4}  + 9\,y_{1}\,y_{4}  + 9\,y_{2}\,y_{3} + 7\,y_{1}\,y_{3} + 5\,y_{1}\,y_{2} - 5\,y_{4} - 4\,y_{3}  - 3\,y_{2} - 2\,y_{1} + 1\big)(1-y_0).\nonumber
\eeqa
A stabilization phenomena occurs such that $\mathcal{C}_k(0)=\mathcal{C}_{k+1}(0)\vert_{y_k=0}$.
A similar construction applies to all scaled cyclic cumulants $g^{(P)}_\sigma(x_1,\cdots,x_P)$. See \cite{Bernard2021Solution} for more details.

However, this construction leaves a few questions open. What is the algebraic structure underlying it ? Is it related to any known structure in integrable systems ? Can it be made more explicit to give a better control on all cumulants for all single cycle permutations ? As pointed out above, the scalings of these cumulants with the system size is such that it ensures the (formal) existence of a large deviation function. Can they be resumed to yield access to this large deviation function ? Does this large deviation function derives from an extremization problem ? Etc. 

\medskip

$\bullet$ {\it Fluctuation dynamics, entanglement statistics and dynamics}

The time evolution of fluctuations is governed by the stochastic equation \eqref{eq:Q-flow}. Clearly, it can be also recasted as a spin chain dynamics, as in eq.\eqref{eq:L-tensorC-GL} for the closed case, once applied to any number of replicas. Since the boundary terms are quadratic in the fermions, they will add boundary terms to eq.\eqref{eq:L-tensorC-GL} linear in the $gl(2R)$ generators \eqref{eq:G-relation}.


For one replica, this spin chain has been shown to be integrable \cite{Essler2020Integrability}. However, as for the closed case, it is still an open question whether it is integrable for any number of replicas.

The scaling limit of the quadratic fluctuations can be studied as in eq.\eqref{eq:ep-g-expand} for the closed case. Clearly, their bulk evolution equations \eqref{eq:g1-evolution} will be unchanged by the boundary processes, and the extra boundary terms in the discrete SDE \eqref{eq:dG-open} will lead to extra boundary contact terms completing these bulk evolution equations. 


As for the closed case, it is important to get a good understanding of entropy statistics and dynamics, either for entanglement entropies or for mutual information of sub-systems relative to their complements. These questions are yet unexplored. In particular, describing accurately the steady statistics of those entropies requires having a good control on the invariant measure.

\medskip

$\bullet$ {\it Un-reasonable connexions with combinatorics}

As it is apparent in the explicit examples, say in eqs.\eqref{eq:C-exemple}, all scaled cumulants $g^{(P)}_\sigma$ (w.r.t. the invariant measure $\mathbb{E}_\infty$) are polynomials with integer coefficients. This is remarkable, but it lacks an a priori explanation.

Since the cumulants involve integer numbers, one may wonder whether they have a combinatorial meaning. And indeed they do. Let us specialize them to coincident points, say $x_k=-t$, for all $k$. Then, we have,
\[ \omega_{n+1}\vert_{\{x_k=-t\}} = - t\, \Phi_n(t)\, (1+t),\]
with $\Phi_n(t)$ polynomials of degree $n-1$ and positive integer coefficients. For instance, cf. eq.\eqref{eq:C-exemple}, 
\begin{eqnarray}
\Phi_2(t) &=& 1 + 2t, \nonumber\\
\Phi_3(t) &=& 1 + 5t+5t^2, \label{eq:count}\\
\Phi_4(t) &=& 1 + 9t+21t^2+14t^3, \nonumber\\
\Phi_5(t) &=& 1 + 14t+56t^2+84t^3+42t^4. \nonumber
\end{eqnarray}
These polynomials are known to count the $(n-k)$ dimensional faces in the associahedron of order $n$ \cite{Devadoss2011discrete}. This is indeed surprising. There is yet no good understanding of this connection except that both structures are related to the moduli spaces of configurations of particles on the line. The observation that the recursion relation \eqref{eq:Ck-series} produces the generating functions of the associahedron has been recently proved \cite{Biane2021UnPublished}. However, the scaled cumulants $g^{(P)}_\sigma$, without evaluating them at coincident points, yields refinements of the associahedron generating functions (indexed by single cycle permutations), and it remains an open question to understand whether their coefficients have natural combinatorial interpretations.

\section{Conclusion}

We have reviewed recent on-going progresses in studies of Quantum Simple Exclusion Processes and their possible connexions with a putative extension of the Macroscopic Fluctuation Theory to the quantum regime, called the {\it Quantum Mesoscopic Fluctuation Theory}. These processes have a rich, interesting but intriguing, structure which is yet to be fully understood. Many questions about the lessons these quantum processes may teach us, or about their structure, remain open. Some of these questions have been formulated all along the review, at the end of each Sections or sub-points they refer to.

\appendix

\section{Appendix~: Does ETH imply ergodicity of quantum trajectories~?}
\label{app:conjecture}

Checking linear ergodicity, as written in eq.\eqref{eq:ergodicity1}, $\lim_{T\to\infty} {T}^{-1} \int_0^T \! ds\, \Tr(\rho_s\, O) = \Tr(\rho_\infty\, O)$, requires information on the system density matrix $\rho_s$ for all time $s\in[0,T]$. If not continuously monitoring the system, this requires preparing it, observing it, and reconstructing its state, at sufficiently different times to sample the time interval $[0,T]$. This may be not exactly what is done is practice...

The situation is better if one is monitoring the quantum system, because recursively extracting information on the system allows to estimate its state and hence the linear \eqref{eq:ergodicity1} or non-linear  \eqref{eq:ergodicity2bis} time averages. It is also more physical because one expects to be able to observe regularly extended systems, without disturbing them too much, and having ergodicity for generic enough thermodynamical systems. 

Monitoring a quantum system is however at the prize of back-acting on it and, as a consequence, its time evolution becomes stochastic. The time evolution of a monitored quantum system is governed by a stochastic differential equation whose solutions are called quantum trajectories, see e.g. \cite{wiseman_milburn_2009,Jacobs2014Book}. For simplicity, for so-called homodyne detections of an observable $O$, the evolution equation is
\beq \label{eq:Q-trajec}
d\rho_t = -i[H,\rho_t]\,dt + \eta\mathcal{L}_O(\rho_t)\, dt + \sqrt{\eta}\,\big(O\rho_t+\rho_tO-2\rho_t\Tr(O\rho_t)\big)\,dB_t ~,
\eeq
with $\mathcal{L}_O(\rho_t)= -\frac{1}{2}[O,[O,\rho_t]]$ and $H$ the hamiltonian of the system in absence of measurement\footnote{One may also add an intrinsic Lindbladian if the system is open.}. Here, $\eta$ is the rate at which information is extracted and  $B_t$ is a Brownian motion related to the output signal $Y_t$ obtained from monitoring the system~:
\beq
dY_t = 2\sqrt{\eta}\,\Tr(O\rho_t) + dB_t ~.
\eeq
Reading this signal yields an estimation of the instantaneous values of the quantum expectation $\Tr(O\rho_t)$. One may extend this equation to cases with multiple monitored observables, say $O^{(a)}$. Monitoring enough observables allow to estimate the density matrix at any time.
Not recording the output signals of the measurements amounts to average the quantum trajectories. The average density matrix $\bar \rho_t$ then satisfies the following dissipative Lindblad equation~: $\partial_t\bar\rho_t = -i[H,\bar\rho_t] + \eta\mathcal{L}_O(\bar\rho_t)$.

Eq.\eqref{eq:Q-trajec} is a well-posed SDE on density matrices. Its linear ergodicity property has been established sometime ago \cite{Kuemmerer2004APE,Kuemmerer2001AnET}. Assuming the uniqueness of the average steady state $\bar \rho_\infty$, solution of $\eta\mathcal{L}_O(\bar \rho_\infty)=i[H,\bar \rho_\infty]$ or its generalization with multiple observables, then  \cite{Kuemmerer2004APE,Kuemmerer2001AnET}
\beq \label{eq:KM-converge}
\lim_{T\to\infty} {T}^{-1} \int_0^T \! ds\, \Tr(\rho_s\, O) = \Tr(\bar \rho_\infty\, O) ~,\quad \mathrm{almost\ surely}.
\eeq
The almost sure convergence here means that eq.\eqref{eq:KM-converge} holds for all output records, except maybe for a set of measure zero.

The non-linear ergodicity \eqref{eq:ergodicity2bis} is then a well-posed problem for the processes \eqref{eq:Q-trajec}. Not so many results about the ergodicity of the quantum trajectories are known, except, of course, when the invariant measure of the SDE  \eqref{eq:Q-trajec} is known to be unique. However, the only known (at least to the author) rigorous results about constructing this invariant measure are those of \cite{benoist2017invariant,benoist2020invariant}. The latter were proved under assumptions whose physical interpretation is not very transparent, in particular in the case of many-body extended systems.

On the other hand, the Eigenstate Thermalisation Hypothesis (ETH) \cite{Deutsch1991Quantum,Srednicki1994Chaos,Srednicki1999Approach} has recently been taken as a sign, if not as a definition, of ergodicity in many-body physics. ETH asserts that the matrix elements of any say local observable ${O}$, in the energy eigenbasis, take the following form : 
 \beq \label{eq:ETH}
 O_{ij}=\mathcal{D}_O({E}_{ij})\delta_{ij}+\sigma(E_{ij})^{-\frac{1}{2}} f_O({E}_{ij},\omega_{ij})R^O_{ij} ,
 \eeq
 with $\sigma(E)$ the density of energy eigenstate and ${E}_{ij}=\frac{1}{2}({E}_i+{E}_j)$ and $\omega_{ij}=\frac{1}{2}({E}_i-{E}_j)$, with $E_i$ the eigen-energies. Here, $\mathcal{D}_O$ and $f_O$ are assumed to be smooth functions, fastly decreasing with $\omega$, and $R^O_{ij}$ matrices of order one with erratically varying elements in the range of energy around ${E}$, with zero mean and unit variance,
  \[ \overline{R^O_{ij}} =0,\quad  \overline{R^O_{ij}R^O_{kl}}=\delta_{jk}\delta_{il} ~.\]

Schematically, this means that the matrix $O_{ij}$ can approximatively be thought of as a band matrix. Its width is governed by the decay rate of $f_O({E},\omega)$ as a function of $\omega$ and it is estimated to be of the order of the temperature $1/\beta_E$ at the energy $E$. The function $f_O({E},\omega)$ is expected to be approximatively constant on an energy scale of the order the Thouless energy $\mathcal{E}_T$, which is the inverse of the diffusion time, $\mathcal{E}_T:={\hbar D}/{L^2}$, with $D$ the diffusion constant and $L$ the linear size of the system. The ``statistics'' of the off-diagonal elements $R^O_{ij}$, around an energy $E$, are in practice defined by sampling its elements in an energy window around $E$. There is of course some arbitrariness in choosing the size of this window, but the Thouless energy $\mathcal{E}_T$ is a natural choice since $f_O({E},\omega)$ is close to a constant on this scale. Even though $\mathcal{E}_T$ decreases with the volume size, there is still an exponentially large number of eigen-energy in this window since eigen-energies are exponentially close (so there is enough energy eigen-states to accurately do the statistics). It is known that higher moments of the off-diagonal elements have to be non-trivial \cite{Foini2019Eigenstate}.

By construction \cite{Deutsch1991Quantum,Srednicki1994Chaos,Srednicki1999Approach}, ETH ensures linear ergodicity. Indeed, suppose that the system is prepared in a pure state in the micro-canonical energy window $E$ up to $\delta E$, say
\[ |\psi\rangle = \sum_i c_i\, |E_i\rangle,\quad \sum_i |c_i|^2 =1 ~,\]
with coefficients $c_i$, smooth and non-vanishing only in the energy window $\delta E$ around the energy $E$. The time evolved state is $|\psi(t)\rangle = \sum_i c_i\,e^{-iE_it}\, |E_i\rangle$ and the quantum expectation value of an observable $O$ is $\langle O \rangle_t=\langle \psi(t)|O|\psi(t)\rangle$ or $\langle O \rangle_t=\sum_{i,j} O_{ij}\, c_ic_j^*e^{i(E_j-E_i)t}$. It is then clear that time averaging this expectation value projects it on its diagonal matrix elements, which are smooth by the ETH, so that \cite{Srednicki1999Approach}
\beq 
 \lim_{T\to \infty} \frac{1}{T} \int_0^T dt \langle O \rangle_t = \sum_i O_{ii} |c_i|^2 \simeq \Tr(\rho_\text{micro}^{(E)}\, O) ~,
 \eeq
 with $\rho_\text{micro}^{(E)}$ the micro-canonical Gibbs state at energy $E$. Thus, assuming ETH, linear ergodicity w.r.t. to the micro-canonical ensemble holds.
 
 Assuming that ETH implies ergodicity of extended many-body systems, as it is often done in the physics literature, it is hence natural to wonder whether (or to conjecture that) ETH ensures ergodicity -- say as formulated in eq.\eqref{eq:ergodicity2bis} --  of the quantum trajectories \eqref{eq:Q-trajec}, or their generalisation with multiple monitored observables, for a large enough set of monitored observables.

\vskip 1.0 truecm

\noindent {\bf Acknowledgements:}\\
I thank my collaborators on this topics, Michel Bauer, Fabian Essler, Ludwig Hruza, Tony Jin, Alexandre Krajenbrink, Marko Medenjak and Lorenzo Piroli, for enjoyable collaborations and Jean-Bernard Zuber for regular discussions.
This work was in part supported by CNRS, by the ENS and by the ANR project ``ESQuisses'', contract number ANR-20-CE47-0014-01. 

\bibliographystyle{ieeetr} 


\bibliography{references}

\end{document}